%% file: pcdecode.tex
\documentclass[journal,12pt,onecolumn]{IEEEtran}
\pdfoutput=1

\usepackage[utf8]{inputenc}
\usepackage[english]{babel}

\usepackage{cite}
\usepackage[cmex10]{amsmath} % Use the [cmex10] option to ensure complicance
\usepackage{amssymb}
\usepackage{bbm}
\usepackage{algorithm}
\usepackage{algpseudocode}
\usepackage{graphicx}
\usepackage{cleveref}

\newcommand{\F}{\ensuremath{\mathbb{F}}}
\newcommand{\Z}{\ensuremath{\mathbb{Z}}}

\newtheorem{definition}{Definition}
\newtheorem{proposition}{Proposition}
\newtheorem{theorem}{Theorem}

\newtheorem{corollary}{Corollary}
\newtheorem{remark}{Remark}

\makeatletter
\newcommand{\removelatexerror}{\let\@latex@error\@gobble}
\makeatother

\begin{document}

\title{
  On Hard-Decision Decoding of Product Codes
}

\author{Ferdinand~Blomqvist
	\thanks{The author is at the Department of Mathematics and Systems Analysis, Aalto University,Helsinki, Finland}
}

\maketitle

\begin{abstract}
  In this paper we review existing hard-decision decoding algorithms for product codes along with different post-processing techniques used in conjunction with the iterative decoder for product codes.
  We improve the decoder by Reddy and Robinson and use it to create a new post-processing technique.
  The performance of this new post-processing technique is evaluated through simulations, and these suggest that our new post-processing technique outperforms previously known post-processing techniques which are not tailored for specific codes.
  The cost of using the new post-processing technique is that the algorithm becomes more complex.
  However, the post-processing is applied very rarely unless the channel is very noisy, and hence the increase in computational complexity is negligible for most choices of parameters.
  Finally, we propose a new algorithm that combines existing techniques in a way that avoids the error floor with short relatively high rate codes.
  The algorithm should also avoid the error floor with long high rate codes, but further work is needed to confirm this.
\end{abstract}


% \fi
\IEEEpeerreviewmaketitle

\section{Introduction}
Product codes form a class of concatenated codes and they were introduced in 1954 by Elias~\cite{elias1954error}.
Hard-decision decoding of product codes is usually carried out with a so-called iterative decoder, which is efficient and can correct most error patterns up to half the minimum distance, and many error patterns beyond half the minimum distance.
The iterative decoder was hinted at by Elias in \cite{elias1954error}, but it was first properly described in \cite{abramson1968cascade}.

The performance of the iterative decoder at low frame error rate (FER), is limited by the occurrence of error patterns of weight less than half the minimum distance that the decoder cannot correct.
These error patterns are usually called \emph{stall patterns}.
The performance of the iterative decoder can be improved with the use of so-called \emph{post-processing} techniques which essentially deal with stall patterns as they are encountered.
Different post-processing techniques are considered in \cite{kreshchuk2014iterative, condo2016stall, emmadi2015half, mittelholzer2016improving}.
Some of the techniques work for any product code while others are limited to certain codes.

There are decoding algorithms for product codes that can correct all errors up to half the minimum distance~\cite{reddy1972random, weldon1971decoding}, but they cannot correct as many error patterns of weight at least half the minimum distance as the iterative decoder mentioned above.
Wainberg \cite{wainberg1972error} extended the decoder from \cite{reddy1972random} to an errors and erasure decoder.

Decoding algorithms for concatenated codes are presented in \cite{blokh1982linear,zinoviev1978gcc,zinoviev1979bursts} and these can be applied to product codes.
These algorithms can also correct all error patters of weight less than half the minimum distance.
The novelty of these algorithms is that they can work for any concatenated code.
However, when applied to product codes they essentially reduce to the algorithm by Reddy and Robinson \cite{reddy1972random}.\footnote{
  \cite{blokh1982linear,zinoviev1978gcc, zinoviev1979bursts} are all in Russian and we have not been able to find translations.
  Our perception of the content in these references is entirely based upon second hand sources such as \cite{kreshchuk2014iterative, ericson1986simple, zyablov1999introduction}.
}

The main idea of all these algorithms that are capable of maximum error correction is the same and it can be traced back to Forney's generalized minimum distance decoding \cite{forney1965concatenated}.

Soft decision decoding of product codes is considered in \cite{pyndiah1998near} and \cite{le2008reed}.

In this paper we review existing post-processing techniques, present an improvement to the decoding algorithm proposed in \cite{reddy1972random}, and use this improved algorithm as part of a new post-processing technique.
We compare the new post-processing technique to other known techniques and evaluate the performance of these different techniques with simulations.
We also propose a new algorithm that is targeted towards high rate product codes.
This algorithm is a simple combination of existing decoders and heuristic arguments imply that it should lower the error floor considerably at low enough FER/BER.

The paper is organized as follows.
Section \ref{sec:prelim} establishes the notation and presents the necessary preliminaries.
In Section \ref{sec:improvegmd} we improve the decoder presented in \cite{reddy1972random},
Section \ref{sec:newpp} is devoted to our new post-processing technique,
and in Section \ref{sec:performance} we present the simulation results.
Section \ref{sec:lowfer} shortly deals with the challenges of avoiding the error floor for very low FER/BER.

\section{Preliminaries and notation}
\label{sec:prelim}
For $n \in \Z$, define $[n] := \{i \in \Z\ | 1 \leq i \leq n \}$, and let $\F_q$ denote the finite field with $q$ elements.
For $c \in \F_{q}^n$, $w(c)$ denotes the Hamming weight of $c$.
Given a vector $x \in \F_{q}^n$, let $x_i$ denote the $i$-th coordinate of $x$.
For two vectors $x,y$, let $x \cdot y$ denote the scalar product of $x$ and $y$.
Let $\mathcal{P}(S)$ denote the power set of $S$.

\begin{definition}
  Let $C$ and $C'$ be linear codes over $\F_{q}$ with parameters $[n,k,d]$ and $[n',k',d']$, and generator matrices $G$ and $G'$, respectively.
  The \emph{product code} $C \times C'$ is the image of the map
  \begin{equation*}
    \sigma: \F_q^{k \times k'} \longrightarrow \F_q^{n \times n'}, \quad
    X \mapsto G^T X G'.
  \end{equation*}
\end{definition}
The following result is well known (see, for instance, \cite[p. 566-567]{macwilliams1977theory}).
\begin{proposition}
  $C \times C'$ is an $[n \cdot n', k \cdot k', d \cdot d']$ code.
\end{proposition}
\begin{remark}
  Product codes are usually defined assuming that the generator matrices of the codes are in systematic form.
  Although it is quite common to assume systematic form of the component codes of a product code, we do not need to make this assumption here.
\end{remark}

\subsection{Generalized minimum distance decoding}
Generalized minimum distance (GMD) decoding was introduced in 1965 by Forney \cite{forney1965concatenated}.
The idea of GMD decoding is to leverage symbol reliability information -- which could be provided, \emph{e.g.}, by the demodulator -- when decoding the received word.

For the remaining part of this section, let $C \subseteq \F_{q}^{n}$ be a code with minimum distance $d$, $c \in C$ and $r \in \F_{q}^{n}$.
Furthermore, let $S(r) := \{ i \in [n]\ |\ r_i \neq 0 \}$ denote the support of $r$, and for $E \subseteq [n]$, define $w_{E}(r) := |S(r) \setminus E|$.
Thus $w_{E}(r)$ is the Hamming weight of $r$ punctured in the coordinates of $E$.
Define
\begin{equation*}
  f_i(r,c) := (-1)^{w(r_i - c_i)}
\end{equation*}
and $f(r,c) := (f_1(r,c), \dots, f_n(r,c))$.
Furthermore, let $\alpha \in [0,1]^{n}$ be a so-called {\em reliability weight vector\/}.
This means that $\alpha_i$ is the reliability of $r_i$.
Note that a smaller reliability weight means that the symbol is considered less reliable.
We then have the following theorem.
\begin{theorem}[Forney \cite{forney1965concatenated}]
  \label{thm:forney1}
  Given $C, r$ and $\alpha$ there is at most one codeword $c \in C$ such that
  \begin{equation}
    \label{eq:gmd}
    \alpha \cdot f(r,c) > n - d.
  \end{equation}
\end{theorem}

Suppose we have $J$ reliability classes with corresponding reliability weights $a_j$, $j \in [J]$, and $a_j \leq a_k$ for $j < k$.
Each symbol is put into one of these reliability classes.
Let $E_0(\alpha) := \emptyset$ and $E_j(\alpha) := \{ i \in [n]\ |\ \alpha_i \leq a_j \}$, $j \in [J]$.
We will omit the $\alpha$ when there is no risk of confusion.
Given an error-and-erasure decoder for $C$, it will decode $r$ (using $E_j$ as the erasures) correctly if $2 w_{E_j}(r-c) + |E_j| < d$.

\begin{theorem}[Forney \cite{forney1965concatenated}]
  If $\alpha \cdot f(r,c) > n - d$, then there exists $0 \leq j \leq J$ such that $2 w_{E_j}(r-c) + |E_j| < d$.
  \label{thm:forney2}
\end{theorem}
Theorem \ref{thm:forney2} shows that GMD decoding can be implemented with an error-and-erasure decoder and Theorem \ref{thm:forney1} can be used to check if the chosen $j$ was correct.

By a \emph{trial}, we denote the act of running the decoder for $C$ with one erasure set and then checking if Theorem \ref{thm:forney1} holds for the decoded word.
Forney noted that:
\begin{corollary}[Forney]
  \label{cor:trials}
  At most $\lfloor (d+1) / 2 \rfloor$ trials are required to decode any received word $r$ that satisfies \eqref{eq:gmd}.
\end{corollary}

Furthermore, since $E_J(\alpha) = [n]$ for all $\alpha$ we do not need to try to decode with the erasure set $E_J$.
Therefore, at most $\min\{J, \lfloor (d+1) / 2 \rfloor\}$ trials are required to decode any received word $r$ that satisfies \eqref{eq:gmd}.

The reasoning behind Corollary \ref{cor:trials} is that, if we erase unreliable symbols one by one (instead of erasing all symbols in one class simultaneously), then the error correction capability will increase by one for every 2 erasures.
This does, however, not directly give us a simple way to skip unnecessary $j$ once we know $\alpha$.
To this end, we finish this section with a theorem which does exactly that.
In addition, it gives us an alternative proof of Corollary \ref{cor:trials}.

Recall that an error-and-erasure decoder that decodes up to the minimum (Elias) distance can be described by a map
\begin{equation*}
  \sigma_{C}: \F_q^n \times \mathcal{P}([n]) \to C \cup \{ \mathcal{B} \}
\end{equation*}
where $(r, E)$ is mapped to the (unique) closest codeword, or to $\mathcal{B}$ if a decoding failure occurs.
Hence $\sigma_{C}(r, E) = c$ if and only if $2 w_{E}(r-c) + |E| < d$.
\begin{theorem}
  Let $F_1 \subset F_2 \subset [n]$, and $|F_2| = |F_1| + 1$.
  If $d - |F_1|$ is even and $\sigma_{C}(r, F_1) \neq \mathcal{B}$, then $\sigma_{C}(r, F_1) = \sigma_{C}(r, F_2)$.
  \label{thm:evenearsure}
  \begin{IEEEproof}
    We prove this by contradiction.
    Let $x_j = \sigma_{C}(r, F_j), j \in \{1,2\}$, and suppose that $x_1 \neq x_2$.
    We have $2 w_{F_1}(r-x_1)+ |F_1| < d$ and $2 w_{F_2}(r-x_1) + |F_2| \geq d$, which implies $w_{F_2}(r-x_1) = w_{F_1}(r-x_1)$.
    It follows that $2 w_{F_1}(r-x_1) + |F_1| = d - 1$, which contradicts the assumption that $d - |F_1|$ is even.
  \end{IEEEproof}
\end{theorem}

\subsection{Iterative decoding of product codes}
\label{sec:decodingalgs}
Let $C$ and $C'$ be linear codes over $\F_{q}$ with parameters $[n,k,d]$ and $[n',k',d']$, and consider the product code $C \times C'$.
We will use this notation for the rest of the article.

Iterative (hard-decision) decoding of product codes was hinted at in Elias original paper \cite{elias1954error}, but it was not properly described.
Abramson proposed the iterative decoder in \cite{abramson1968cascade}.
Our description of the iterative decoder is based on the one in \cite{kreshchuk2014iterative}.

Let $r$ denote the received word.
The standard iterative decoder works as follows:
\begin{enumerate}
  \item Decode all columns of $r$ with the column code and correct all errors and erasures.
    Denote the result by $r'$.
  \item Decode all rows of $r'$ with the row code and correct all errors and erasures, and denote the result by $x$.
  \item If $x \neq r$, set $r := x$ and repeat from Step 1.
    Otherwise, go to Step 4.
  \item If there were any decoding failures or errors that were corrected during the last invocations of steps 1 and 2, return failure.
    Otherwise, return $x$.
\end{enumerate}
This iterative decoder is efficient and performs well.
It can correct many error patterns with weight beyond half of the minimum distance, but unfortunately, it cannot correct all error patterns below half of the minimum distance.
There are also variations of the iterative decoder that perform at most $n$ iterations, where $n$ is a constant which is usually quite small, say $2$ or $4$.

The error patterns that cannot be corrected by the iterative decoder are often called \emph{stall patterns} of \emph{stoping sets}.
We will use the former terminology.
These stall patterns limit the performance of the iterative decoder at lower FER.
To combat this, so called \emph{post-processing} techniques have been developed to improve the error floor.

Kreshchuk \emph{et al.} noticed that, when the iterative decoder fails, it corrects all errors outside of an error submatrix and then stops.
However, if one would insert erasures, the decoding process could be continued.
Therefore, they proposed the following decoder \cite{kreshchuk2014iterative}.
\begin{enumerate}
  \item Run the iterative decoder described above.
    Return its results if it succeeds, otherwise continue with Step 2.
  \item Denote all rows that changed or were rejected by the row code during the last iteration by $E_r$.
    Similarly denote the corresponding columns by $E_c$.
  \item Let $E = \{ (r, c)\ |\ r \in E_r, c \in E_c \}$.
    Take the last word produced by the iterative decoder and insert erasures at every position found in $E$.
    Denote this word by $u$.
  \item Run the iterative decoder with $u$ as input.
  \item If the decoder succeeds, return its result.
    Otherwise, reject this word and return a failure.
\end{enumerate}
This decoder can be seen as a post-processing technique; first the iterative decoder is run and whenever it fails the post-processing is applied to the result of the iterative decoder.

A similar post-processing technique is proposed in \cite{emmadi2015half}.
Whenever the iterative decoder fails, mark any rows where a decoding failure occurred as erased.
Then decode this word with a slightly modified iterative decoder; whenever there is a row or column decoding failure mark the corresponding row or column as erased.

Another post-processing technique is proposed in \cite{condo2016stall} by Condo \emph{et al.}.
Their technique is only directly applicable to \emph{extended}-polynomial codes, \emph{i.e.}, polynomial codes with an additional parity symbol added.
To describe their method we introduce some notation.
Suppose that the iterative decoder cannot correct the given word.
Denote all rows were we had a decoding failure during the last iteration by $E_r$, and similarly denote the corresponding columns by $E_c$.
Let $E = \{ (r, c)\ |\ r \in E_r, c \in E_c \}$.
Their post-processing technique for binary codes is simple; flip the bits in every position found in $E$, and run the iterative decoder on the result.

Whenever the code is non-binary the symbol at the intersection of a row and column has many bits so simply flipping these bits is not particularly useful.
Instead Condo \emph{et al.} use the extra parity symbols of the extended codes to determine which bits should be flipped.
The iterative decoder is then run on the word that results from the bit flipping.
More precisely, Condo {et al.} prescribe one iteration of iterative decoding after the bit flipping for both the binary and non-binary case.

Since this post-processing technique needs the extra parity symbols of the extended codes, it is not possible to use it as such for product codes with general non-binary codes.
One can, however, use a slightly simplified version where every symbol at positions found in $E$ are erased.
This gives a post-processing technique that is very similar to the one proposed by Kreshchuk \emph{et al.}.
In fact the only difference is in the definition of $E$; Kreshchuk \emph{et al.} also include rows and columns that changed during the last iteration, while Condo \emph{et al.} only include rows and columns were the decoding failed.

There are other post-processing algorithms for product code like codes such as \emph{half-product} codes and \emph{braided} codes.
We will not consider these techniques, but the interested reader can find further information in, for instance, \cite{mittelholzer2016improving, jian2013iterative}.

\subsection{Other decoding algorithms for product codes}
In \cite{reddy1972random}, Reddy and Robinson presented a decoding algorithm for product codes that can correct all error patterns up to half of the minimum distance.
There are also other algorithms that are capable of maximal error correction, most notably \cite{weldon1971decoding, blokh1982linear}, but we will not discuss these due to the similarity to the algorithm proposed by Reddy and Robinson.

In order to describe the algorithm, we need to introduce some notation.
Let $y = x + e$ be the received word, and let $\hat{x}$  denote the result after decoding every column of $y$ with the column code $C$ and let $\hat{e}$ denote the corresponding error matrix.
Given a word $b$ (as an $n\times n'$-matrix), let $b_j$, and $b^j$ denote the $j$-th column and row of $b$ respectively.
Finally, let $t = \lfloor (d-1) / 2 \rfloor$.

The algorithm proceeds as follows:
\begin{enumerate}
  \item Decode all the columns with the column code $C$ to obtain $\hat{x}$ and $\hat{e}$.
  \item Assign reliability weights $\alpha_j$ to each column by letting
    \begin{equation*}
      \alpha_j :=
      \begin{cases}
	\frac{d - 2 w(\hat{e}_j)}{d},
	  & \text{if } w(\hat{e}_j) \leq t, \\
	0, & \text{if } w(\hat{e}_j) > t \text{ or decoding failure.}
      \end{cases}
    \end{equation*}
  \item Decode every row of $\hat{x}$ with a generalized minimum distance decoder for the row code $C'$ (using $\alpha$ as the reliability weight vector).
\end{enumerate}
This algorithm can correct every error pattern such that the generalized minimum distance criterion \eqref{eq:gmd} holds for all rows after the column decoding.
Therefore, as proved in the original paper, the algorithm can correct all error patterns of weight less than half the minimum distance.
It can also correct many error patterns of larger weight, but apparently the iterative decoder can correct more error patterns.

The well informed reader might notice that the reliability weights defined here are slightly different from the ones given by Reddy and Robinson.
We have chosen to employ the weights as defined by Wainberg \cite{wainberg1972error}, reduced to the case of no erasures, since this way the weighing scheme is more coherent.

Wainberg provides a proof of the correctness of his weighing scheme in \cite{wainberg1972burst}, which seems hard to come by.
Therefore -- and for completeness -- we include a proof of correctness for the algorithm.

The idea of the proof is simple; Suppose that $w(e) < d \cdot d'$ and show that~\eqref{eq:gmd} holds for every row after the column decoding.
We start by estimating the number of errors in every column.
First, suppose that the column decoder decodes correctly, \emph{i.e.}, $\hat{x}_i = x_i$.
Then
\begin{equation}
  \label{eq:correctrow}
  w(e_i) = w(\hat{e}_i) = \frac{d - d + 2 w(\hat{e}_i)}{2} = \frac{1 - \alpha_i}{2} d.
\end{equation}
If, on the other hand, the $i$-th column is incorrectly decoded to another word, then
\begin{equation}
  \label{eq:wrongrow}
  w(e_i) \geq d - w(\hat{e}_i) = \frac{d + d - 2 w(\hat{e}_i)}{2} = \frac{1 + \alpha_i}{2} d.
\end{equation}
Finally, if we have a decoding failure, then $\alpha_i = 0$, and hence $w(e_i) \geq d/2 = (1 + \alpha_i) d / 2$.

Let $I_C$ and $I_E$ be the index sets of the columns that were correctly and erroneously decoded respectively.
$I_E$ also contains the indices of columns where a decoding failure occurred.
Clearly
\begin{equation*}
  \sum_{i \in I_C} \frac{1 - \alpha_i}{2} d
  + \sum_{i \in I_E} \frac{1 + \alpha_i}{2} d
  \leq w(e) < \frac{d \cdot d'}{2},
\end{equation*}
and thus
\begin{equation*}
  \sum_{i \in I_C} (1 - \alpha_i)
  + \sum_{i \in I_E} (1 + \alpha_i)
  = n - \sum_{i \in I_C} \alpha_i
  + \sum_{i \in I_E} \alpha_i
  < d'.
\end{equation*}
Letting
\begin{equation*}
  \varphi_i(I_E) := (-1)^{\mathbbm{1}_{I_e}(i)}
\end{equation*}
and $\varphi(I_E) := (\varphi_1(I_E), \dots, \varphi_n(I_E))$ gives
\begin{equation}
  \label{eq:rowforney}
  \alpha \cdot \varphi(I_E) > n - d'.
\end{equation}

This shows that \eqref{eq:gmd} holds for every row of $\hat{x}$.
More precisely, $\alpha \cdot f(\hat{x}^i, x^i) > n - d'$, for $i \in [n]$.
Hence we can conclude that the algorithm decodes all error patterns of weight less than $d \cdot d' / 2$.

This result is satisfactory, but it can easily be improved.
Let
\begin{equation*}
  w_D(e) := \sum_{i=0}^{n'} \min\{w(e_i), d \}.
\end{equation*}
Both \eqref{eq:correctrow} and \eqref{eq:wrongrow} are still valid if we replace the left hand side with $\min\{w(e_i), d \}$, and hence it follows that the algorithm can correct any error pattern that satisfies $2 w_D(e) < d \cdot d'$.

The computational complexity of the algorithm is simple to analyze.
The decoder for the column code is run $n'$ times and the decoder for the row code is run at most $nm$ times, where $m$ is the maximum number of trials the GMD decoder needs to run to recover $x^i$ from $\hat{x}^i$.

Recall that $\min\{J, \lfloor (d' + 1)/ 2 \rfloor\}$ trials suffice, where $J$ is the number of reliability classes.
However, any symbol with reliability weight zero can always be erased.
To see this, suppose that $2 |I_E| < d'$.
Clearly $E_1 \subseteq I_E$, and hence
\begin{equation*}
  2 |I_E \setminus E_1| + |E_1| \leq 2 |I_E| < d'.
\end{equation*}
Since, $w_{E_1}(\hat{x}^i - x^i) \leq |I_E \setminus E_1|$ for all $i \in [n]$, the claim follows.
This means that the GMD decoder needs to run at most $J-1$ trials,

From the definition of the reliability weights we see that we have $J = \lfloor (d-1)/2\rfloor \ + 2$ reliability classes.
Therefore $x^i$ can be recovered in at most
\begin{equation*}
  m = \left\lfloor \frac{\min \{d, d' \} + 1}{2} \right\rfloor
\end{equation*}
trials.

We will end this section with a few observations that are useful when implementing the algorithm.
These observations do not, unfortunately, affect the worst case complexity of the algorithm.
They do, however, provide a way to eliminate unnecessary trials after we know the received word.

If there are no columns with reliability weight $a_j$, then $E_{j} = E_{j-1}$, and hence we do not need to run the trial for this value of $j$.
Furthermore, by Theorem \ref{thm:evenearsure}, if $j$ is such that $d' - |E_j|$ is even and $|E_{j+1}| = |E_j| + 1$, then this trial can be omitted.
We say that $j \in [J-1]$ is \emph{viable} if $j=1$ or if $j$ does not fulfill either of the two previous conditions.
The GMD decoder only needs to run trials with viable $j$.

\section{Improving the Reddy and Robinson algorithm}
\label{sec:improvegmd}
The Reddy and Robinson algorithm can be optimized in a way that lowers the worst case complexity significantly.
As noted in Section \ref{sec:decodingalgs}, after decoding the columns there exists $1 \leq j < J$ such that $2 |I_E \setminus E_j| + |E_j| < d'$.
This means that there exists one $j$ such that every row will be correctly decoded with the erasure set $E_j$.
Hence the GMD decoder does not need to start from $j=1$ for every row.
Instead it can start from the same $j$ that was used when the previous row was correctly decoded.
This way the row decoder needs to be run at most $n + m - 1$ instead of $nm$ times, where $m = \lfloor (\min\{d', d\} + 1) / 2 \rfloor$.
We call this optimized version of the Reddy and Robinson decoder the {\tt gmd} decoder.

To the best of our knowledge this small improvement has not been presented anywhere in previous papers.

The algorithm can also be modified to an algorithm that, according to our simulations, performs significantly better.
The change is as simple as modifying the GMD decoder used for the row decoding slightly; instead of only decoding up to the GMD -- which means only whenever \eqref{eq:gmd} is satisfied -- we choose the word that maximizes the left hand side of~\eqref{eq:gmd}.
More precisely, given a received word $r$ along with a reliability weight vector $\alpha$, the modified decoder operates as follows,
\begin{enumerate}
  \item For all viable $j$, decode $r$ with an error-and-erasure decoder with the erasure set $E_j(\alpha)$, and denote the result by $c_j$.
  \item Return the $c_j$ that maximizes $\alpha \cdot f(r, c_j)$.
\end{enumerate}
We call this variation of the GMD decoder a {\em generalized distance\/} (GD) {\em decoder\/}.
We use the name {\tt gd} to refer to the improved version of the Reddy and Robinson decoder that uses the GD decoder for the row decoding.

\begin{proposition}
  {\tt gd} correctly decodes any word that {\tt gmd} decodes correctly.
  \begin{IEEEproof}
    Suppose that {\tt gmd} decodes $y = x + e$ correctly.
    Then, after the column decoding, every row satisfies \eqref{eq:gmd}.
    Hence, $\alpha \cdot f(\hat{x}^i, x^i) > n - d'$, and since $x^i$ is the unique codeword of $C'$ that satisfies this condition, the GD decoder will also correctly decode the $i$-th row to $x^i$.
  \end{IEEEproof}
\end{proposition}

\section{A new post-processing technique}
\label{sec:newpp}
To improve the iterative decoder one needs to successfully deal with the stall patterns.
There are essentially two options: use a post-processing technique or resort to another decoding algorithm whenever the iterative decoder fails.
In this case the other decoder would be run on the received word.

The {\tt gd} algorithm seems to be a good choice for an algorithm to combine with the iterative decoder; {\tt gd} can correct all the stall patterns with sufficiently low error weight and thus the algorithms complement each other.
This would result in the following algorithm.
Let $r = x + e$ be the received word.
Try to decode $r$ with the iterative decoder. If it succeeds, return the results.
Otherwise, try to decode $r$ with {\tt gd}.

While this seems like a good approach, it turns out that if we apply {\tt gd} as a post-processing technique, then the resulting algorithm performs significantly better than the algorithm outlined above.
Thus we propose the following algorithm.
Let $r = x + e$ be the received word.
Try to decode $r$ with the iterative decoder. If it succeeds, return the results.
Otherwise, let $u$ denote the word where the iterative decoder stalls and try to decode $u$ with {\tt gd}.

Note that, while it is possible that $u = r$, this seems to be the exception rather than the norm (at least with larger symbol error probabilities).
Hence, these two approaches give very different results.

To determine how this new post-processing technique fares, we have chosen to compare it to other known technique with simulations.
We have chosen to compare it to the techniques proposed by Kreshchuk \emph{et al.}, Emmadi \emph{et al.}, and Condo \emph{et al.}.

\section{Performance}
\label{sec:performance}
We have compared four different post-processing techniques:
\begin{enumerate}
  \item The technique by Kreshchuk \emph{et al.} \cite{kreshchuk2014iterative};
  \item The technique proposed by Emmadi \emph{et al.} in \cite{emmadi2015half};
  \item The technique used by Condo \emph{et al.} in \cite{condo2016stall} modified for use with non-extended codes as described in Section \ref{sec:decodingalgs};
  \item The technique proposed in Section \ref{sec:newpp}.
\end{enumerate}
We also compare each of these techniques to the standard iterative decoder, meaning no post-processing at all.
Furthermore, we compare {\tt gd} to {\tt gmd}.

All of these post-processing techniques introduce erasures, and thus the column and row decoders are required to be error-and-erasure decoders.
Such decoders are widely available for Reed-Solomon codes.
Therefore, we have chosen to run the simulations with product codes constructed from Reed-Solomon codes.
The simulations where performed with the {\tt pcdecode} software package \cite{pcdecode2019github} over a $q$-ary symmetric channel.

\subsection{Simulation results}
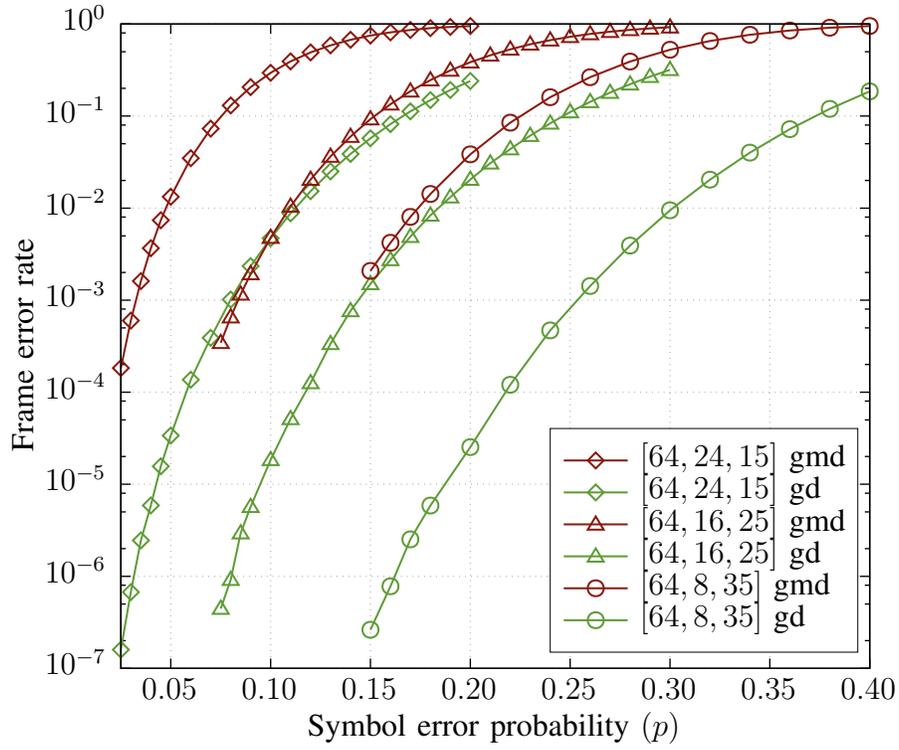
\begin{figure}[htbp]
  \centering
  \scalebox{1.0}{\input{images/sim-gmd-gd.tex}}
  \caption{Simulations with different codes of length $64$.
    The component codes are Reed-Solomon codes of length $8$.}
  \label{fig:gmd-gd}
\end{figure}

The simulations show that {\tt gd} performs significantly better than {\tt gmd}, see Figure \ref{fig:gmd-gd} and \ref{fig:256gmd-gd}.
We have obtained similar results for codes of other lengths and error correcting capabilities.
There is also a very clear pattern between code length, error correcting capability of the code and the gap between {\tt gd} and {\tt gmd}: the longer the code and the more errors it can correct, the bigger the gap between the algorithms.
These results are not surprising, since the difference between the performance achieved with minimum distance decoding is significantly worse than what can be achieved with maximum likelihood decoding.

The simulation results for the different post-processing techniques are presented in Figures \labelcref{fig:64_4-2-4,fig:64_4-4-4,fig:256_5-2-4,fig:256_5-4-4,fig:1024_8-2-4,fig:2304-2-4}.
The results are the same whether we consider the FER or BER.
We have chosen to only show the FER in an effort to make the plots more readable.

\begin{figure}[htbp]
  \centering
  \scalebox{1.0}{\input{images/sim256-gmd-gd.tex}}
  \caption{Simulations with different codes of length $256$.
    The component codes are Reed-Solomon codes of length $16$.}
  \label{fig:256gmd-gd}
\end{figure}
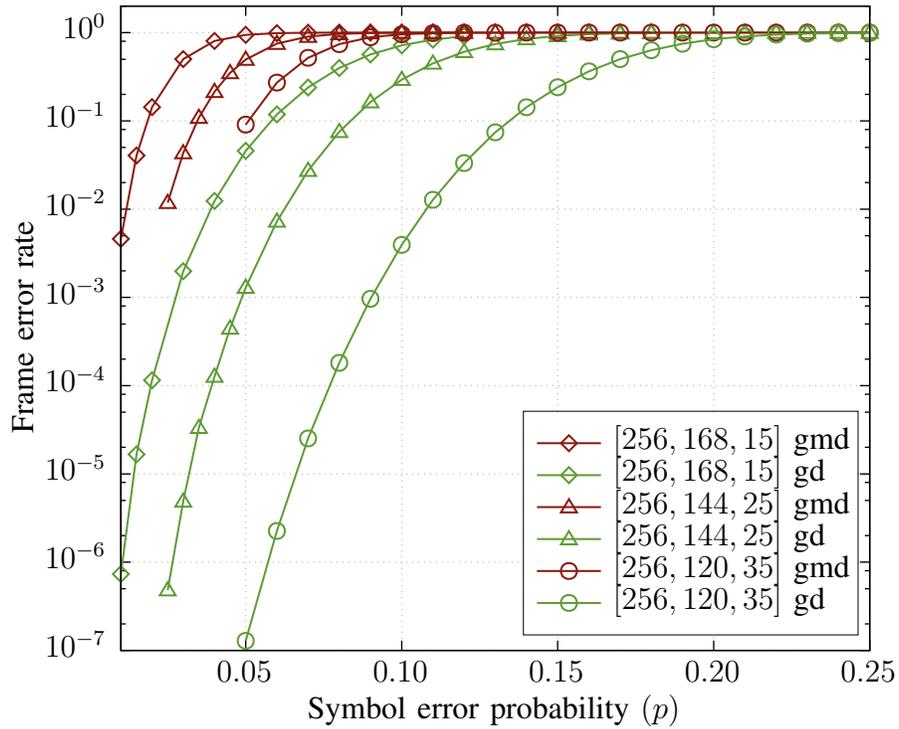
\begin{figure}[htbp]
  \centering
  \scalebox{1.0}{\input{images/sim64_4-2-4.tex}}
  \caption{Simulation with a $[64,24,15]_{16}$ code that is the product of $[8,4,5]_{16}$ and $[8,6,3]_{16}$ Reed-Solomon codes.}
  \label{fig:64_4-2-4}
\end{figure}
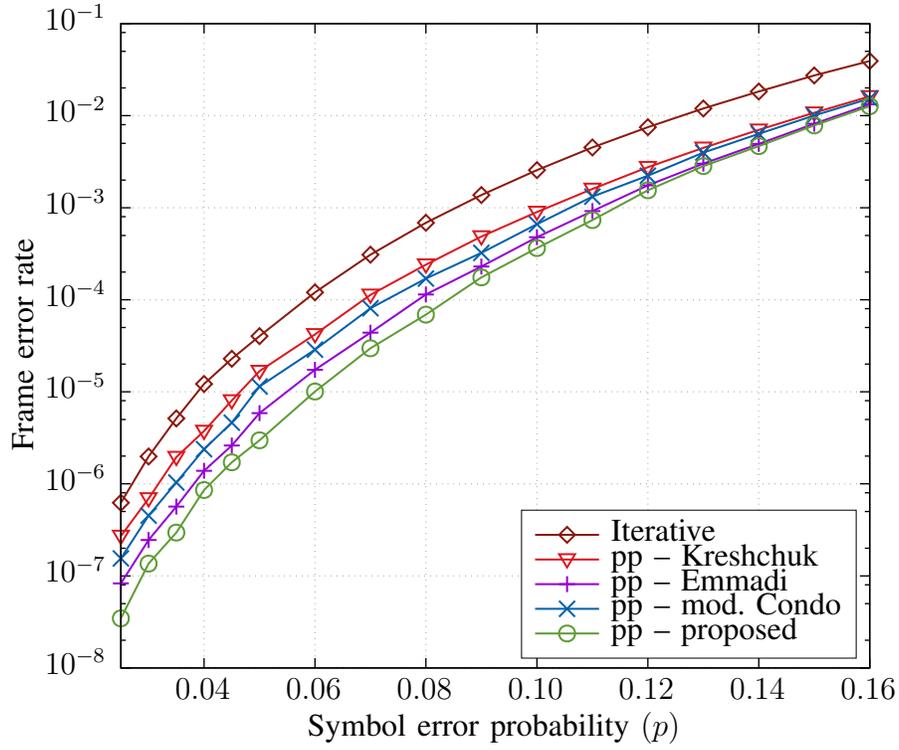
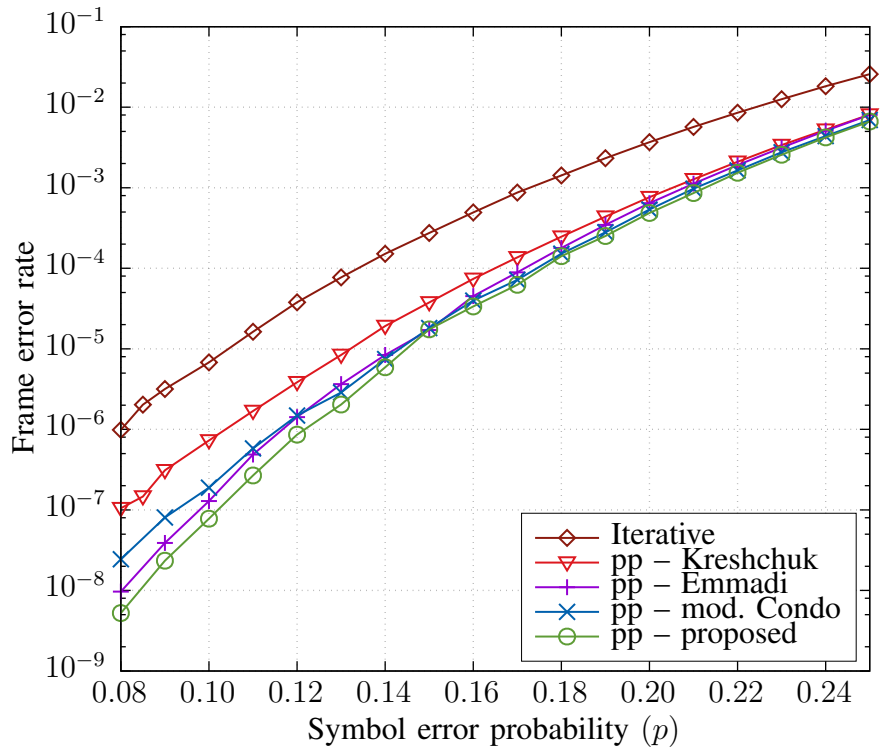
\begin{figure}[htbp]
  \centering
  \scalebox{1.0}{\input{images/sim64_4-4-4.tex}}
  \caption{Simulation with a $[64,16,25]_{16}$ code that is the product of $[8,4,5]_{16}$ and $[8,4,5]_{16}$ Reed-Solomon codes.}
  \label{fig:64_4-4-4}
\end{figure}
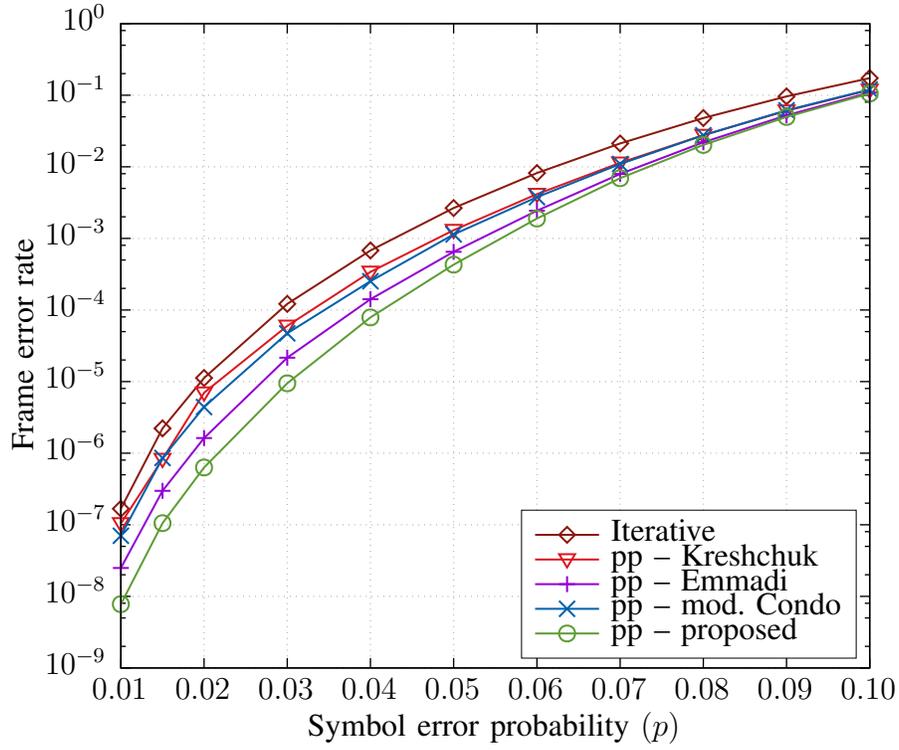
\begin{figure}[htbp]
  \centering
  \scalebox{1.0}{\input{images/sim256_5-2-4.tex}}
  \caption{Simulation with a $[256,168,15]_{32}$ code that is the product of $[16,12,5]_{32}$ and $[16,14,3]_{32}$ Reed-Solomon codes.}
  \label{fig:256_5-2-4}
\end{figure}
\begin{figure}[htbp]
  \centering
  \scalebox{1.0}{\input{images/sim256_5-4-4.tex}}
  \caption{Simulation with a $[256,144,25]_{32}$ code that is the product of $[16,12,5]_{32}$ and $[16,12,5]_{32}$ Reed-Solomon codes.}
  \label{fig:256_5-4-4}
\end{figure}
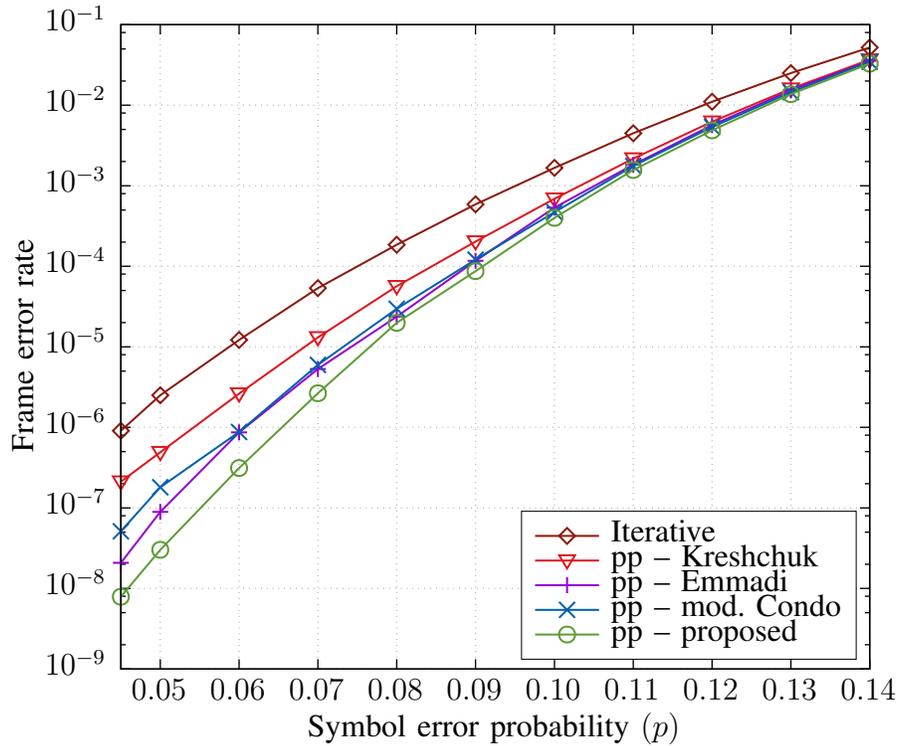
\begin{figure}[htbp]
  \centering
  \scalebox{1.0}{\input{images/sim1024_8-2-4.tex}}
  \caption{Simulation with a $[1024,840,15]_{256}$ code that is the product of $[32,28,5]_{256}$ and $[32,30,3]_{256}$ Reed-Solomon codes.}
  \label{fig:1024_8-2-4}
\end{figure}
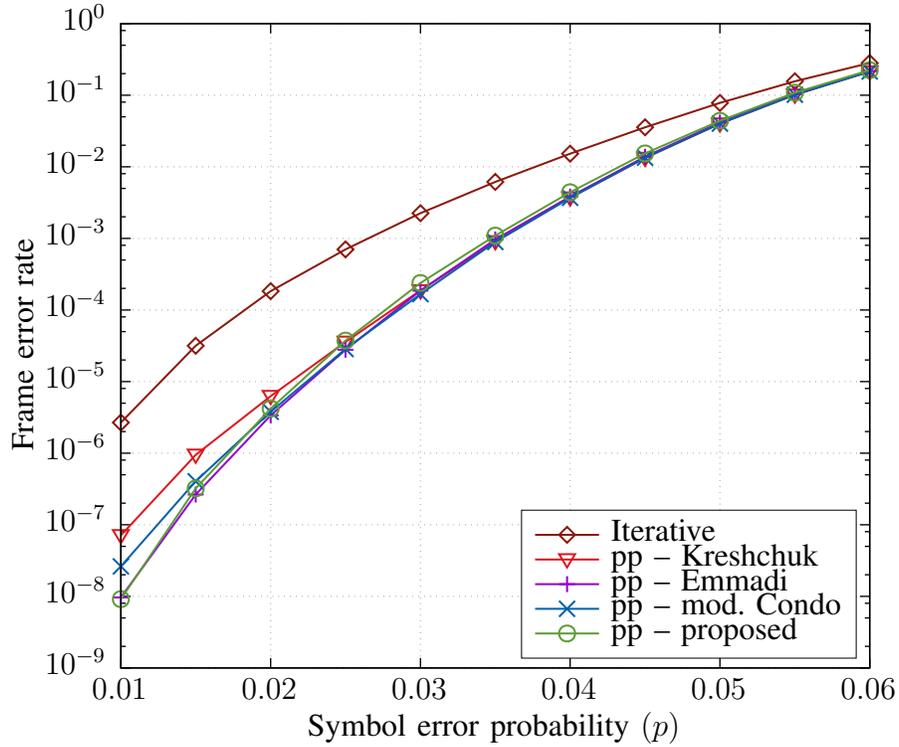
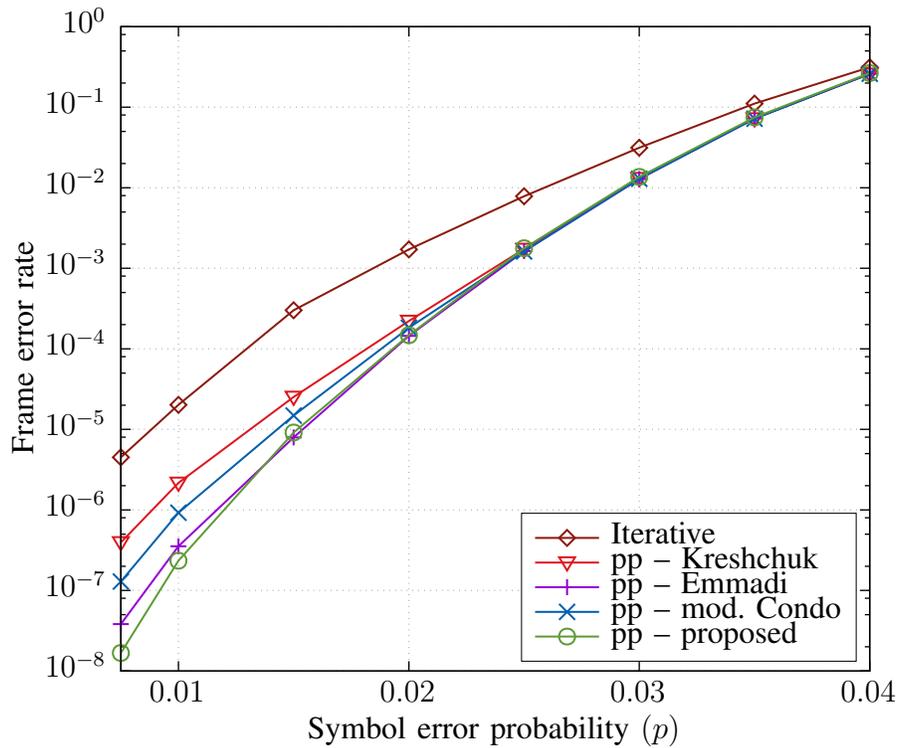
\begin{figure}[htbp]
  \centering
  \scalebox{1.0}{\input{images/sim2304-2-4.tex}}
  \caption{Simulation with a $[2304,2024,15]_{256}$ code that is the product of $[48,44,5]_{256}$ and $[48,46,3]_{256}$ Reed-Solomon codes.}
  \label{fig:2304-2-4}
\end{figure}
We see that the post-processing techniques can be roughly ordered as follows.
The technique by Kreshchuk \emph{et al.}, performs better than the iterative decoder, but worse than all others.
The modified version of the technique by Condo \emph{et al.} performs slightly better, but not much.
Emmadi's method outperforms both of the former techniques, while our proposed technique gives the lowest error rates.
We say that this is a rough order since there is slight variability to this depending on the specific code used.

\subsection{Column-first vs. row-first decoding}
All of the decoding algorithms reviewed here start by decoding the columns.
Due to the symmetric properties of product codes, we could as well have chosen to use the row-first versions of the algorithms.
This question is only relevant whenever the row and column code have different error correcting capabilities.

The question of column-first versus row-first decoding can equivalently be stated as: if we have component codes with differing minimum distance, which one should we choose as the column code.
Here we assume that columns-first decoding is used.

Simulations suggest the following results.
The {\tt gmd} decoder performs better if one decodes the less powerful code first, while {\tt gd} and the iterative decoder perform better if the more powerful code is decoded first.

It seems reasonable to use the more powerful code first, and thus the results for the iterative decoder and {\tt gd} are hardly surprising.
A short explanation for {\tt gmd} behaving differently is as follows.

Suppose that we have codes $C$ and $C'$ as in Section \ref{sec:prelim} such that $d < d'$.
If we consider the product code $C \times C'$, then {\tt gmd} can correct all error patterns $e$ such that
\begin{equation}
  \label{eq:lesspowerful}
  \sum_{i=1}^{n'} \min\{w(e_i), d \} < \frac{d \cdot d'}{2}.
\end{equation}
On the other hand, if we swap the minimum distances of $C$ and $C'$, then {\tt gmd} can correct all patterns that satisfy
\begin{equation}
  \label{eq:morepowerful}
  \sum_{i=1}^{n'} \min\{w(e_i), d' \} < \frac{d \cdot d'}{2}.
\end{equation}
There are clearly more error patterns $e$ that satisfy \eqref{eq:lesspowerful} then there are patterns that satisfy \eqref{eq:morepowerful}.

\subsection{Notes on computational complexities}
\begin{figure}[htbp]
  \centering
  \scalebox{1.0}{\input{images/gamma.tex}}
  \caption{$\gamma$ as a function of the frame error rate for a variety of codes.}
  \label{fig:gamma}
\end{figure}
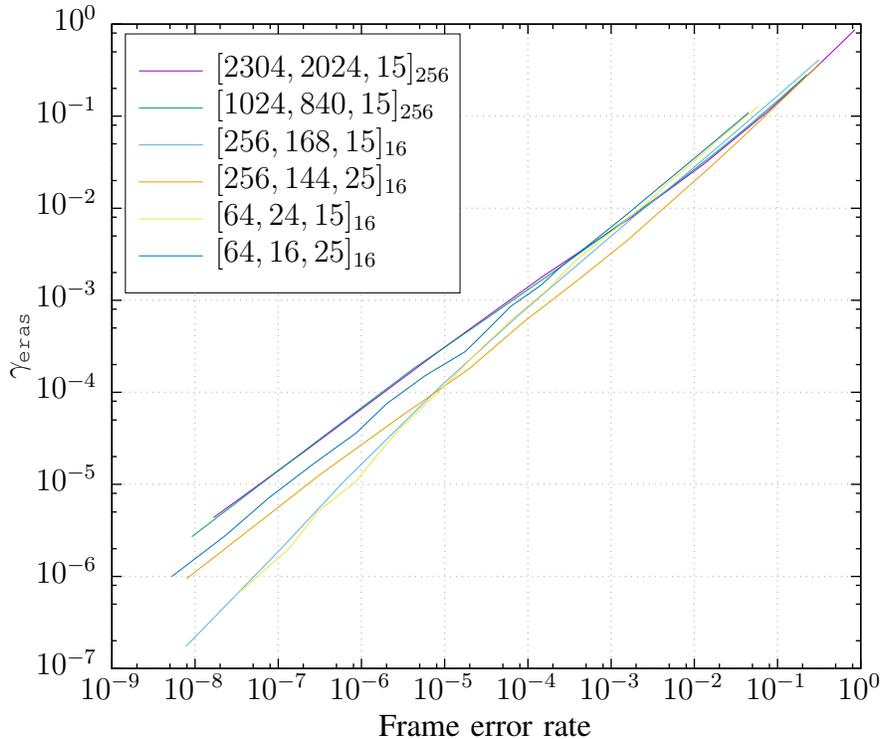

Applying post-processing techniques to the iterative decoder can significantly lower the error rate at medium to low FER/BER.
It does however come at the cost of increased computational complexity.
To analyze the impact of post-processing on the average computational complexity we consider the ratio of the number of times the post-processing was invoked and the total number of words processed.
Denote this ratio by $\gamma$.
Figure \ref{fig:gamma} show how $\gamma$ depends on the FER for a variety of different codes.
The ratios are computed using data gathered from the simulations with our proposed post-processing technique.
However, $\gamma$ only depends on the properties of the iterative decoder, and hence the results are the same for all post-processing techniques.
We can clearly see that, for any reasonable channel quality, the post-processing is applied very seldom.
Therefore, there is no reason not to apply post-processing (at least not for the majority of applications).
Furthermore, the computational complexity of the post-processing technique does not matter unless it is orders of magnitude larger than that of the iterative decoder.

The computational complexities of all the post-processing technique reviewed are similar and hence one should choose the one that gives the lowest FER/BER.

\section{Reaching very low FER/BER}
\label{sec:lowfer}
Simulations with short and relatively high rate codes show that the iterative decoder along with post-processing techniques still exhibits an error floor at quite high FER.
The underlying problem is that, even with post-processing, the decoder cannot correct all error patterns with weight below half of the minimum distance of the code.
Although we cannot verify it with simulations, the same error floor -- albeit at a much lower FER -- should also be present with longer high rate codes if the symbol error probability is low enough.
One potential solution to this problem is to combine the {\tt gmd} decoder with the iterative decoder plus post-processing in the following way.
First run the {\tt gmd} decoder on the received word.
If it succeeds, then stop and return the decoded word.
Otherwise, run the iterative decoder with post-processing on the received word.

This decoder can correct all error patterns of weight less than half the minimum distance and also all the error patterns that the iterative decoder with post-processing can correct.\footnote{
  Here we opportunistically assuming that the {\tt gmd} decoder does not misscorrect any received word.
The probability of this happening is very low when the symbol error probability is low enough.
}
Hence the error floor will be considerably lower.

Simulations with short codes (length 36, 64 and 100) suggest that this algorithm fares well.
More precisely, these simulations suggest that the error rate of the algorithm is upper bounded by the iterative decoder with post processing, and that the gap between the algorithms increases when the symbol error probability is lowered.
Any meaningful improvement in FER is only achieved with very low symbol error probabilities.

For long codes the error floor of the iterative decoder with post-processing is only reached at a FER that is so low that it is unfeasible to run reliable simulations at this FER.
Therefore other techniques must be applied to validate the performance of the algorithm for very low FER.

It is also possible to apply this algorithm to half-product codes.
Further work in this direction and comparisons to techniques used for half product codes, for instance those in \cite{mittelholzer2016improving}, will be left for future work.

\section{Conlusions}
In this paper, we have studied and compared different hard-decision decoding algorithms for product codes.
We have improved the Reddy and Robinson decoder~\cite{reddy1972random} by allowing decoding beyond the generalized minimum distance.
Furthermore, we have presented a new post-processing technique that utilizes this improved decoder.
Our simulations suggest that our new post-processing technique outperforms other known post-processing techniques that are not optimized for specific codes.

In addition, a new algorithm that is targeted towards high rate codes is presented.
Heuristic arguments are in favor for the algorithm but further work is needed to give useful upper bounds for the FER achievable with this method.

\bibliographystyle{IEEEtran}
\bibliography{pcdecode}

\end{document}

%% file: images/sim-gmd-gd.tex
% GNUPLOT: LaTeX picture with Postscript
\begingroup
  \makeatletter
  \providecommand\color[2][]{%
    \GenericError{(gnuplot) \space\space\space\@spaces}{%
      Package color not loaded in conjunction with
      terminal option `colourtext'%
    }{See the gnuplot documentation for explanation.%
    }{Either use 'blacktext' in gnuplot or load the package
      color.sty in LaTeX.}%
    \renewcommand\color[2][]{}%
  }%
  \providecommand\includegraphics[2][]{%
    \GenericError{(gnuplot) \space\space\space\@spaces}{%
      Package graphicx or graphics not loaded%
    }{See the gnuplot documentation for explanation.%
    }{The gnuplot epslatex terminal needs graphicx.sty or graphics.sty.}%
    \renewcommand\includegraphics[2][]{}%
  }%
  \providecommand\rotatebox[2]{#2}%
  \@ifundefined{ifGPcolor}{%
    \newif\ifGPcolor
    \GPcolortrue
  }{}%
  \@ifundefined{ifGPblacktext}{%
    \newif\ifGPblacktext
    \GPblacktexttrue
  }{}%
  % define a \g@addto@macro without @ in the name:
  \let\gplgaddtomacro\g@addto@macro
  % define empty templates for all commands taking text:
  \gdef\gplbacktext{}%
  \gdef\gplfronttext{}%
  \makeatother
  \ifGPblacktext
    % no textcolor at all
    \def\colorrgb#1{}%
    \def\colorgray#1{}%
  \else
    % gray or color?
    \ifGPcolor
      \def\colorrgb#1{\color[rgb]{#1}}%
      \def\colorgray#1{\color[gray]{#1}}%
      \expandafter\def\csname LTw\endcsname{\color{white}}%
      \expandafter\def\csname LTb\endcsname{\color{black}}%
      \expandafter\def\csname LTa\endcsname{\color{black}}%
      \expandafter\def\csname LT0\endcsname{\color[rgb]{1,0,0}}%
      \expandafter\def\csname LT1\endcsname{\color[rgb]{0,1,0}}%
      \expandafter\def\csname LT2\endcsname{\color[rgb]{0,0,1}}%
      \expandafter\def\csname LT3\endcsname{\color[rgb]{1,0,1}}%
      \expandafter\def\csname LT4\endcsname{\color[rgb]{0,1,1}}%
      \expandafter\def\csname LT5\endcsname{\color[rgb]{1,1,0}}%
      \expandafter\def\csname LT6\endcsname{\color[rgb]{0,0,0}}%
      \expandafter\def\csname LT7\endcsname{\color[rgb]{1,0.3,0}}%
      \expandafter\def\csname LT8\endcsname{\color[rgb]{0.5,0.5,0.5}}%
    \else
      % gray
      \def\colorrgb#1{\color{black}}%
      \def\colorgray#1{\color[gray]{#1}}%
      \expandafter\def\csname LTw\endcsname{\color{white}}%
      \expandafter\def\csname LTb\endcsname{\color{black}}%
      \expandafter\def\csname LTa\endcsname{\color{black}}%
      \expandafter\def\csname LT0\endcsname{\color{black}}%
      \expandafter\def\csname LT1\endcsname{\color{black}}%
      \expandafter\def\csname LT2\endcsname{\color{black}}%
      \expandafter\def\csname LT3\endcsname{\color{black}}%
      \expandafter\def\csname LT4\endcsname{\color{black}}%
      \expandafter\def\csname LT5\endcsname{\color{black}}%
      \expandafter\def\csname LT6\endcsname{\color{black}}%
      \expandafter\def\csname LT7\endcsname{\color{black}}%
      \expandafter\def\csname LT8\endcsname{\color{black}}%
    \fi
  \fi
    \setlength{\unitlength}{0.0500bp}%
    \ifx\gptboxheight\undefined%
      \newlength{\gptboxheight}%
      \newlength{\gptboxwidth}%
      \newsavebox{\gptboxtext}%
    \fi%
    \setlength{\fboxrule}{0.5pt}%
    \setlength{\fboxsep}{1pt}%
\begin{picture}(6800.00,5660.00)%
    \gplgaddtomacro\gplbacktext{%
      \csname LTb\endcsname%
      \put(747,595){\makebox(0,0)[r]{\strut{}$10^{-7}$}}%
      \csname LTb\endcsname%
      \put(747,1289){\makebox(0,0)[r]{\strut{}$10^{-6}$}}%
      \csname LTb\endcsname%
      \put(747,1984){\makebox(0,0)[r]{\strut{}$10^{-5}$}}%
      \csname LTb\endcsname%
      \put(747,2678){\makebox(0,0)[r]{\strut{}$10^{-4}$}}%
      \csname LTb\endcsname%
      \put(747,3372){\makebox(0,0)[r]{\strut{}$10^{-3}$}}%
      \csname LTb\endcsname%
      \put(747,4066){\makebox(0,0)[r]{\strut{}$10^{-2}$}}%
      \csname LTb\endcsname%
      \put(747,4761){\makebox(0,0)[r]{\strut{}$10^{-1}$}}%
      \csname LTb\endcsname%
      \put(747,5455){\makebox(0,0)[r]{\strut{}$10^{0}$}}%
      \csname LTb\endcsname%
      \put(1225,409){\makebox(0,0){\strut{}$0.05$}}%
      \csname LTb\endcsname%
      \put(1978,409){\makebox(0,0){\strut{}$0.10$}}%
      \csname LTb\endcsname%
      \put(2730,409){\makebox(0,0){\strut{}$0.15$}}%
      \csname LTb\endcsname%
      \put(3483,409){\makebox(0,0){\strut{}$0.20$}}%
      \csname LTb\endcsname%
      \put(4235,409){\makebox(0,0){\strut{}$0.25$}}%
      \csname LTb\endcsname%
      \put(4988,409){\makebox(0,0){\strut{}$0.30$}}%
      \csname LTb\endcsname%
      \put(5740,409){\makebox(0,0){\strut{}$0.35$}}%
      \csname LTb\endcsname%
      \put(6493,409){\makebox(0,0){\strut{}$0.40$}}%
    }%
    \gplgaddtomacro\gplfronttext{%
      \csname LTb\endcsname%
      \put(144,3025){\rotatebox{-270}{\makebox(0,0){\strut{}Frame error rate}}}%
      \csname LTb\endcsname%
      \put(3671,130){\makebox(0,0){\strut{}Symbol error probability $(p)$}}%
      \csname LTb\endcsname%
      \put(4771,2163){\makebox(0,0)[l]{\strut{}$[64,24,15]$ gmd}}%
      \csname LTb\endcsname%
      \put(4771,1922){\makebox(0,0)[l]{\strut{}$[64,24,15]$ gd}}%
      \csname LTb\endcsname%
      \put(4771,1681){\makebox(0,0)[l]{\strut{}$[64,16,25]$ gmd}}%
      \csname LTb\endcsname%
      \put(4771,1440){\makebox(0,0)[l]{\strut{}$[64,16,25]$ gd}}%
      \csname LTb\endcsname%
      \put(4771,1199){\makebox(0,0)[l]{\strut{}$[64,8,35]$ gmd}}%
      \csname LTb\endcsname%
      \put(4771,958){\makebox(0,0)[l]{\strut{}$[64,8,35]$ gd}}%
    }%
    \gplbacktext
    \put(0,0){\includegraphics{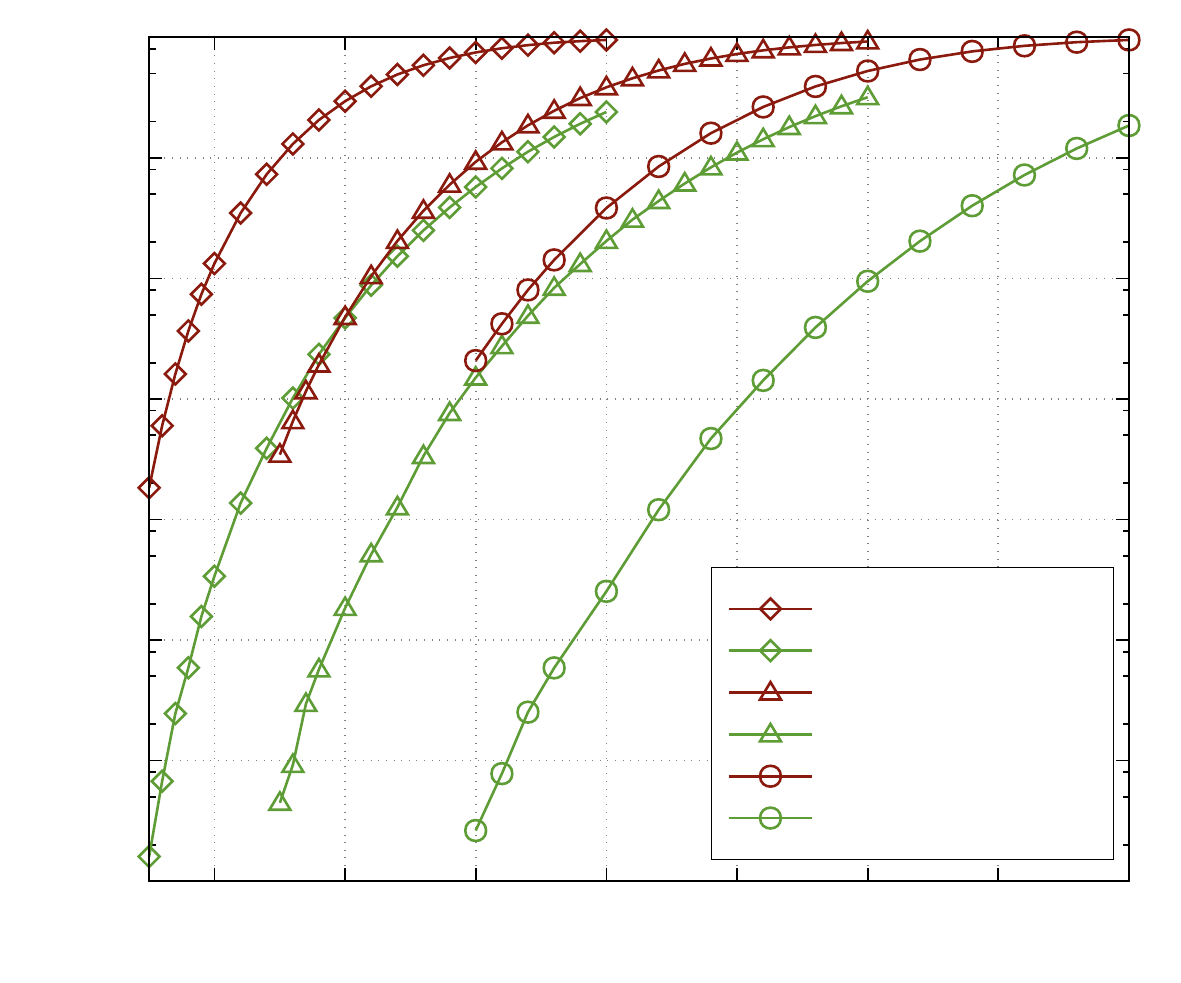}}%
    \gplfronttext
  \end{picture}%
\endgroup

%% file: images/sim256-gmd-gd.tex
% GNUPLOT: LaTeX picture with Postscript
\begingroup
  \makeatletter
  \providecommand\color[2][]{%
    \GenericError{(gnuplot) \space\space\space\@spaces}{%
      Package color not loaded in conjunction with
      terminal option `colourtext'%
    }{See the gnuplot documentation for explanation.%
    }{Either use 'blacktext' in gnuplot or load the package
      color.sty in LaTeX.}%
    \renewcommand\color[2][]{}%
  }%
  \providecommand\includegraphics[2][]{%
    \GenericError{(gnuplot) \space\space\space\@spaces}{%
      Package graphicx or graphics not loaded%
    }{See the gnuplot documentation for explanation.%
    }{The gnuplot epslatex terminal needs graphicx.sty or graphics.sty.}%
    \renewcommand\includegraphics[2][]{}%
  }%
  \providecommand\rotatebox[2]{#2}%
  \@ifundefined{ifGPcolor}{%
    \newif\ifGPcolor
    \GPcolortrue
  }{}%
  \@ifundefined{ifGPblacktext}{%
    \newif\ifGPblacktext
    \GPblacktexttrue
  }{}%
  % define a \g@addto@macro without @ in the name:
  \let\gplgaddtomacro\g@addto@macro
  % define empty templates for all commands taking text:
  \gdef\gplbacktext{}%
  \gdef\gplfronttext{}%
  \makeatother
  \ifGPblacktext
    % no textcolor at all
    \def\colorrgb#1{}%
    \def\colorgray#1{}%
  \else
    % gray or color?
    \ifGPcolor
      \def\colorrgb#1{\color[rgb]{#1}}%
      \def\colorgray#1{\color[gray]{#1}}%
      \expandafter\def\csname LTw\endcsname{\color{white}}%
      \expandafter\def\csname LTb\endcsname{\color{black}}%
      \expandafter\def\csname LTa\endcsname{\color{black}}%
      \expandafter\def\csname LT0\endcsname{\color[rgb]{1,0,0}}%
      \expandafter\def\csname LT1\endcsname{\color[rgb]{0,1,0}}%
      \expandafter\def\csname LT2\endcsname{\color[rgb]{0,0,1}}%
      \expandafter\def\csname LT3\endcsname{\color[rgb]{1,0,1}}%
      \expandafter\def\csname LT4\endcsname{\color[rgb]{0,1,1}}%
      \expandafter\def\csname LT5\endcsname{\color[rgb]{1,1,0}}%
      \expandafter\def\csname LT6\endcsname{\color[rgb]{0,0,0}}%
      \expandafter\def\csname LT7\endcsname{\color[rgb]{1,0.3,0}}%
      \expandafter\def\csname LT8\endcsname{\color[rgb]{0.5,0.5,0.5}}%
    \else
      % gray
      \def\colorrgb#1{\color{black}}%
      \def\colorgray#1{\color[gray]{#1}}%
      \expandafter\def\csname LTw\endcsname{\color{white}}%
      \expandafter\def\csname LTb\endcsname{\color{black}}%
      \expandafter\def\csname LTa\endcsname{\color{black}}%
      \expandafter\def\csname LT0\endcsname{\color{black}}%
      \expandafter\def\csname LT1\endcsname{\color{black}}%
      \expandafter\def\csname LT2\endcsname{\color{black}}%
      \expandafter\def\csname LT3\endcsname{\color{black}}%
      \expandafter\def\csname LT4\endcsname{\color{black}}%
      \expandafter\def\csname LT5\endcsname{\color{black}}%
      \expandafter\def\csname LT6\endcsname{\color{black}}%
      \expandafter\def\csname LT7\endcsname{\color{black}}%
      \expandafter\def\csname LT8\endcsname{\color{black}}%
    \fi
  \fi
    \setlength{\unitlength}{0.0500bp}%
    \ifx\gptboxheight\undefined%
      \newlength{\gptboxheight}%
      \newlength{\gptboxwidth}%
      \newsavebox{\gptboxtext}%
    \fi%
    \setlength{\fboxrule}{0.5pt}%
    \setlength{\fboxsep}{1pt}%
\begin{picture}(6800.00,5660.00)%
    \gplgaddtomacro\gplbacktext{%
      \csname LTb\endcsname%
      \put(747,595){\makebox(0,0)[r]{\strut{}$10^{-7}$}}%
      \csname LTb\endcsname%
      \put(747,1261){\makebox(0,0)[r]{\strut{}$10^{-6}$}}%
      \csname LTb\endcsname%
      \put(747,1926){\makebox(0,0)[r]{\strut{}$10^{-5}$}}%
      \csname LTb\endcsname%
      \put(747,2592){\makebox(0,0)[r]{\strut{}$10^{-4}$}}%
      \csname LTb\endcsname%
      \put(747,3258){\makebox(0,0)[r]{\strut{}$10^{-3}$}}%
      \csname LTb\endcsname%
      \put(747,3923){\makebox(0,0)[r]{\strut{}$10^{-2}$}}%
      \csname LTb\endcsname%
      \put(747,4589){\makebox(0,0)[r]{\strut{}$10^{-1}$}}%
      \csname LTb\endcsname%
      \put(747,5255){\makebox(0,0)[r]{\strut{}$10^{0}$}}%
      \csname LTb\endcsname%
      \put(1790,409){\makebox(0,0){\strut{}$0.05$}}%
      \csname LTb\endcsname%
      \put(2966,409){\makebox(0,0){\strut{}$0.10$}}%
      \csname LTb\endcsname%
      \put(4141,409){\makebox(0,0){\strut{}$0.15$}}%
      \csname LTb\endcsname%
      \put(5317,409){\makebox(0,0){\strut{}$0.20$}}%
      \csname LTb\endcsname%
      \put(6493,409){\makebox(0,0){\strut{}$0.25$}}%
    }%
    \gplgaddtomacro\gplfronttext{%
      \csname LTb\endcsname%
      \put(144,3025){\rotatebox{-270}{\makebox(0,0){\strut{}Frame error rate}}}%
      \csname LTb\endcsname%
      \put(3671,130){\makebox(0,0){\strut{}Symbol error probability $(p)$}}%
      \csname LTb\endcsname%
      \put(4569,2158){\makebox(0,0)[l]{\strut{}$[256,168,15]$ gmd}}%
      \csname LTb\endcsname%
      \put(4569,1917){\makebox(0,0)[l]{\strut{}$[256,168,15]$ gd}}%
      \csname LTb\endcsname%
      \put(4569,1676){\makebox(0,0)[l]{\strut{}$[256,144,25]$ gmd}}%
      \csname LTb\endcsname%
      \put(4569,1435){\makebox(0,0)[l]{\strut{}$[256,144,25]$ gd}}%
      \csname LTb\endcsname%
      \put(4569,1194){\makebox(0,0)[l]{\strut{}$[256,120,35]$ gmd}}%
      \csname LTb\endcsname%
      \put(4569,953){\makebox(0,0)[l]{\strut{}$[256,120,35]$ gd}}%
    }%
    \gplbacktext
    \put(0,0){\includegraphics{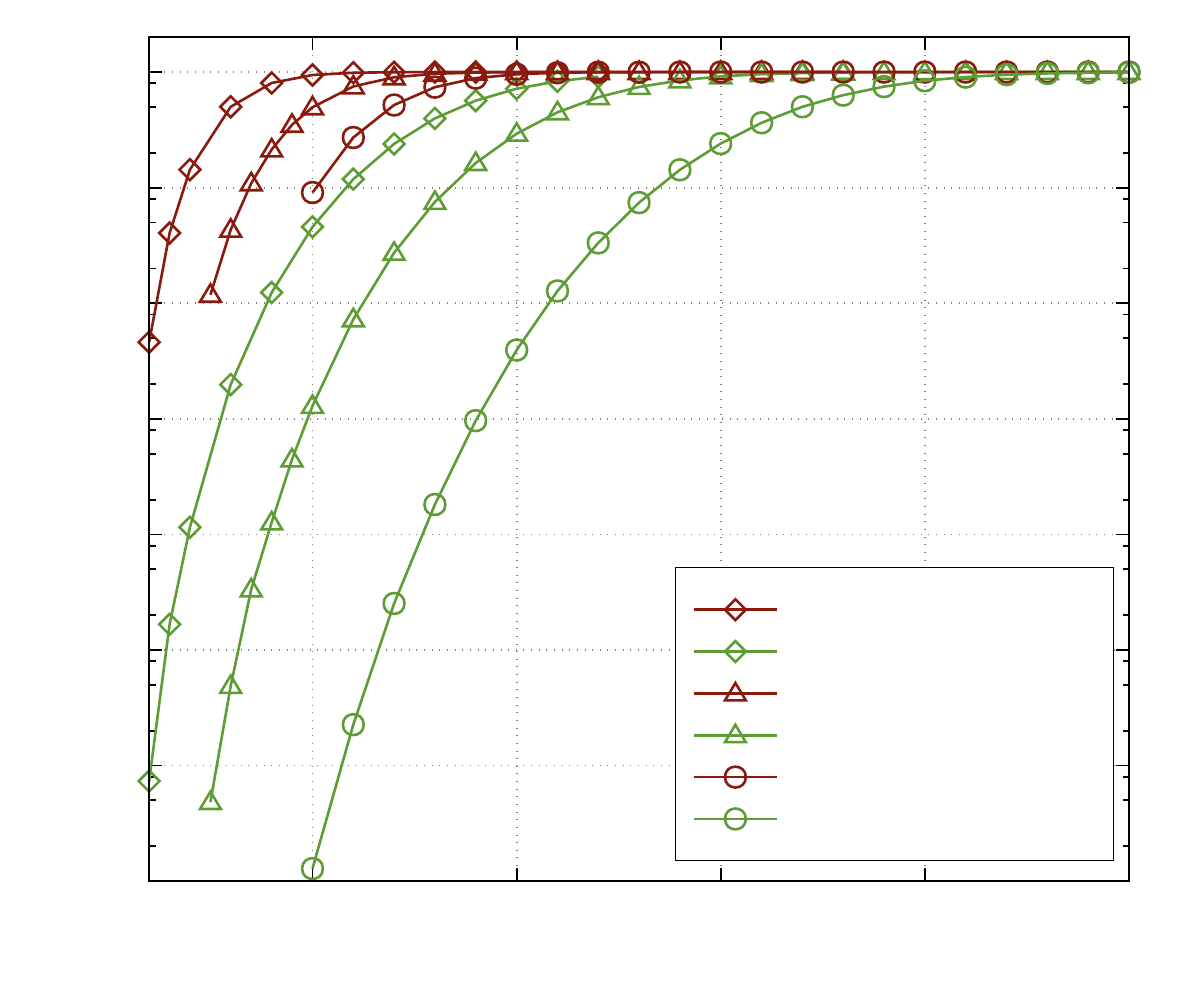}}%
    \gplfronttext
  \end{picture}%
\endgroup

%% file: images/sim64_4-2-4.tex
% GNUPLOT: LaTeX picture with Postscript
\begingroup
  \makeatletter
  \providecommand\color[2][]{%
    \GenericError{(gnuplot) \space\space\space\@spaces}{%
      Package color not loaded in conjunction with
      terminal option `colourtext'%
    }{See the gnuplot documentation for explanation.%
    }{Either use 'blacktext' in gnuplot or load the package
      color.sty in LaTeX.}%
    \renewcommand\color[2][]{}%
  }%
  \providecommand\includegraphics[2][]{%
    \GenericError{(gnuplot) \space\space\space\@spaces}{%
      Package graphicx or graphics not loaded%
    }{See the gnuplot documentation for explanation.%
    }{The gnuplot epslatex terminal needs graphicx.sty or graphics.sty.}%
    \renewcommand\includegraphics[2][]{}%
  }%
  \providecommand\rotatebox[2]{#2}%
  \@ifundefined{ifGPcolor}{%
    \newif\ifGPcolor
    \GPcolortrue
  }{}%
  \@ifundefined{ifGPblacktext}{%
    \newif\ifGPblacktext
    \GPblacktexttrue
  }{}%
  % define a \g@addto@macro without @ in the name:
  \let\gplgaddtomacro\g@addto@macro
  % define empty templates for all commands taking text:
  \gdef\gplbacktext{}%
  \gdef\gplfronttext{}%
  \makeatother
  \ifGPblacktext
    % no textcolor at all
    \def\colorrgb#1{}%
    \def\colorgray#1{}%
  \else
    % gray or color?
    \ifGPcolor
      \def\colorrgb#1{\color[rgb]{#1}}%
      \def\colorgray#1{\color[gray]{#1}}%
      \expandafter\def\csname LTw\endcsname{\color{white}}%
      \expandafter\def\csname LTb\endcsname{\color{black}}%
      \expandafter\def\csname LTa\endcsname{\color{black}}%
      \expandafter\def\csname LT0\endcsname{\color[rgb]{1,0,0}}%
      \expandafter\def\csname LT1\endcsname{\color[rgb]{0,1,0}}%
      \expandafter\def\csname LT2\endcsname{\color[rgb]{0,0,1}}%
      \expandafter\def\csname LT3\endcsname{\color[rgb]{1,0,1}}%
      \expandafter\def\csname LT4\endcsname{\color[rgb]{0,1,1}}%
      \expandafter\def\csname LT5\endcsname{\color[rgb]{1,1,0}}%
      \expandafter\def\csname LT6\endcsname{\color[rgb]{0,0,0}}%
      \expandafter\def\csname LT7\endcsname{\color[rgb]{1,0.3,0}}%
      \expandafter\def\csname LT8\endcsname{\color[rgb]{0.5,0.5,0.5}}%
    \else
      % gray
      \def\colorrgb#1{\color{black}}%
      \def\colorgray#1{\color[gray]{#1}}%
      \expandafter\def\csname LTw\endcsname{\color{white}}%
      \expandafter\def\csname LTb\endcsname{\color{black}}%
      \expandafter\def\csname LTa\endcsname{\color{black}}%
      \expandafter\def\csname LT0\endcsname{\color{black}}%
      \expandafter\def\csname LT1\endcsname{\color{black}}%
      \expandafter\def\csname LT2\endcsname{\color{black}}%
      \expandafter\def\csname LT3\endcsname{\color{black}}%
      \expandafter\def\csname LT4\endcsname{\color{black}}%
      \expandafter\def\csname LT5\endcsname{\color{black}}%
      \expandafter\def\csname LT6\endcsname{\color{black}}%
      \expandafter\def\csname LT7\endcsname{\color{black}}%
      \expandafter\def\csname LT8\endcsname{\color{black}}%
    \fi
  \fi
    \setlength{\unitlength}{0.0500bp}%
    \ifx\gptboxheight\undefined%
      \newlength{\gptboxheight}%
      \newlength{\gptboxwidth}%
      \newsavebox{\gptboxtext}%
    \fi%
    \setlength{\fboxrule}{0.5pt}%
    \setlength{\fboxsep}{1pt}%
\begin{picture}(6800.00,5660.00)%
    \gplgaddtomacro\gplbacktext{%
      \csname LTb\endcsname%
      \put(747,595){\makebox(0,0)[r]{\strut{}$10^{-8}$}}%
      \csname LTb\endcsname%
      \put(747,1289){\makebox(0,0)[r]{\strut{}$10^{-7}$}}%
      \csname LTb\endcsname%
      \put(747,1984){\makebox(0,0)[r]{\strut{}$10^{-6}$}}%
      \csname LTb\endcsname%
      \put(747,2678){\makebox(0,0)[r]{\strut{}$10^{-5}$}}%
      \csname LTb\endcsname%
      \put(747,3372){\makebox(0,0)[r]{\strut{}$10^{-4}$}}%
      \csname LTb\endcsname%
      \put(747,4066){\makebox(0,0)[r]{\strut{}$10^{-3}$}}%
      \csname LTb\endcsname%
      \put(747,4761){\makebox(0,0)[r]{\strut{}$10^{-2}$}}%
      \csname LTb\endcsname%
      \put(747,5455){\makebox(0,0)[r]{\strut{}$10^{-1}$}}%
      \csname LTb\endcsname%
      \put(1476,409){\makebox(0,0){\strut{}$0.04$}}%
      \csname LTb\endcsname%
      \put(2312,409){\makebox(0,0){\strut{}$0.06$}}%
      \csname LTb\endcsname%
      \put(3148,409){\makebox(0,0){\strut{}$0.08$}}%
      \csname LTb\endcsname%
      \put(3985,409){\makebox(0,0){\strut{}$0.10$}}%
      \csname LTb\endcsname%
      \put(4821,409){\makebox(0,0){\strut{}$0.12$}}%
      \csname LTb\endcsname%
      \put(5657,409){\makebox(0,0){\strut{}$0.14$}}%
      \csname LTb\endcsname%
      \put(6493,409){\makebox(0,0){\strut{}$0.16$}}%
    }%
    \gplgaddtomacro\gplfronttext{%
      \csname LTb\endcsname%
      \put(144,3025){\rotatebox{-270}{\makebox(0,0){\strut{}Frame error rate}}}%
      \csname LTb\endcsname%
      \put(3671,130){\makebox(0,0){\strut{}Symbol error probability $(p)$}}%
      \csname LTb\endcsname%
      \put(4555,1599){\makebox(0,0)[l]{\strut{}Iterative}}%
      \csname LTb\endcsname%
      \put(4555,1413){\makebox(0,0)[l]{\strut{}pp -- Kreshchuk}}%
      \csname LTb\endcsname%
      \put(4555,1227){\makebox(0,0)[l]{\strut{}pp -- Emmadi}}%
      \csname LTb\endcsname%
      \put(4555,1041){\makebox(0,0)[l]{\strut{}pp -- mod. Condo}}%
      \csname LTb\endcsname%
      \put(4555,855){\makebox(0,0)[l]{\strut{}pp -- proposed}}%
    }%
    \gplbacktext
    \put(0,0){\includegraphics{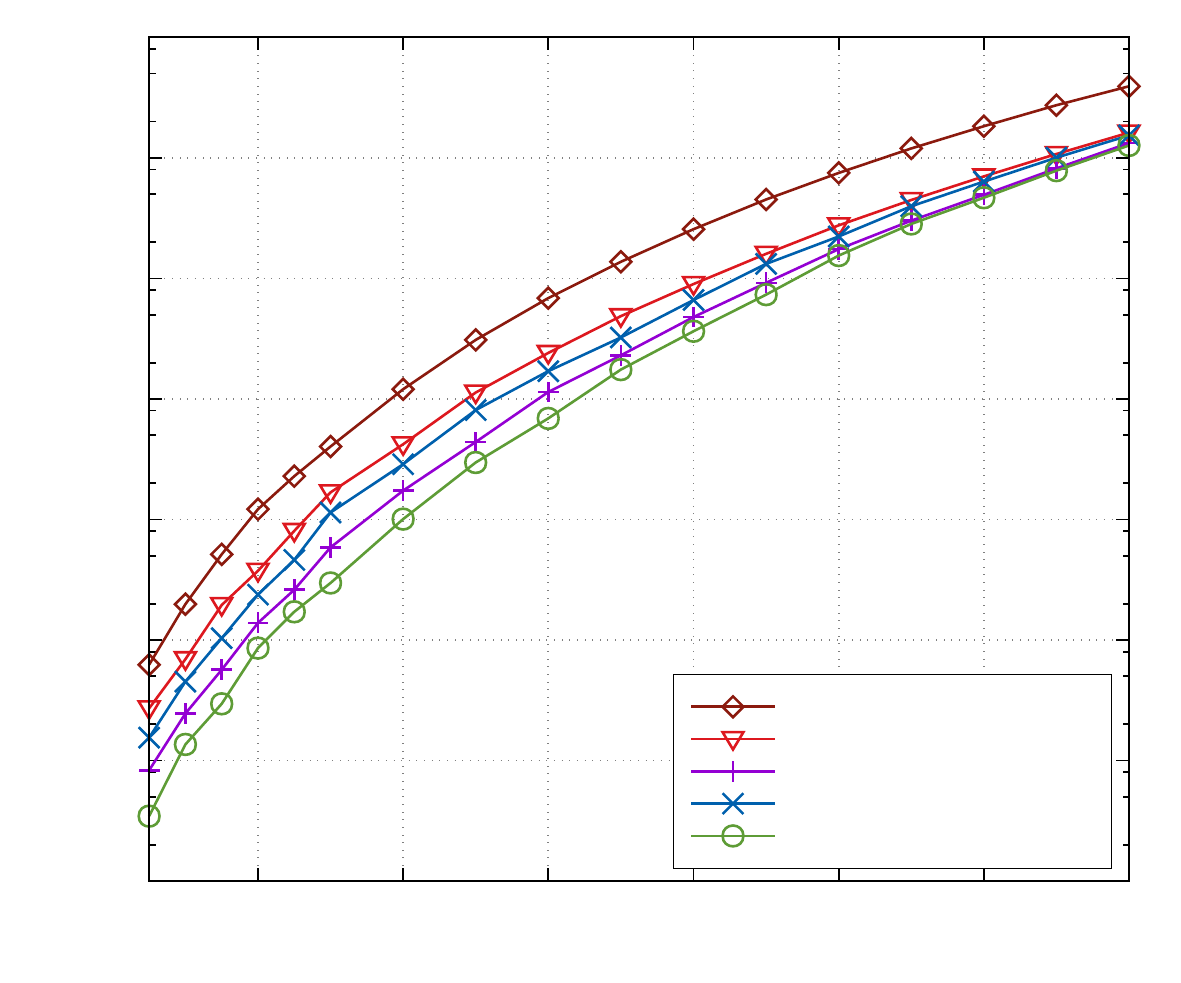}}%
    \gplfronttext
  \end{picture}%
\endgroup

%% file: images/sim64_4-4-4.tex
% GNUPLOT: LaTeX picture with Postscript
\begingroup
  \makeatletter
  \providecommand\color[2][]{%
    \GenericError{(gnuplot) \space\space\space\@spaces}{%
      Package color not loaded in conjunction with
      terminal option `colourtext'%
    }{See the gnuplot documentation for explanation.%
    }{Either use 'blacktext' in gnuplot or load the package
      color.sty in LaTeX.}%
    \renewcommand\color[2][]{}%
  }%
  \providecommand\includegraphics[2][]{%
    \GenericError{(gnuplot) \space\space\space\@spaces}{%
      Package graphicx or graphics not loaded%
    }{See the gnuplot documentation for explanation.%
    }{The gnuplot epslatex terminal needs graphicx.sty or graphics.sty.}%
    \renewcommand\includegraphics[2][]{}%
  }%
  \providecommand\rotatebox[2]{#2}%
  \@ifundefined{ifGPcolor}{%
    \newif\ifGPcolor
    \GPcolortrue
  }{}%
  \@ifundefined{ifGPblacktext}{%
    \newif\ifGPblacktext
    \GPblacktexttrue
  }{}%
  % define a \g@addto@macro without @ in the name:
  \let\gplgaddtomacro\g@addto@macro
  % define empty templates for all commands taking text:
  \gdef\gplbacktext{}%
  \gdef\gplfronttext{}%
  \makeatother
  \ifGPblacktext
    % no textcolor at all
    \def\colorrgb#1{}%
    \def\colorgray#1{}%
  \else
    % gray or color?
    \ifGPcolor
      \def\colorrgb#1{\color[rgb]{#1}}%
      \def\colorgray#1{\color[gray]{#1}}%
      \expandafter\def\csname LTw\endcsname{\color{white}}%
      \expandafter\def\csname LTb\endcsname{\color{black}}%
      \expandafter\def\csname LTa\endcsname{\color{black}}%
      \expandafter\def\csname LT0\endcsname{\color[rgb]{1,0,0}}%
      \expandafter\def\csname LT1\endcsname{\color[rgb]{0,1,0}}%
      \expandafter\def\csname LT2\endcsname{\color[rgb]{0,0,1}}%
      \expandafter\def\csname LT3\endcsname{\color[rgb]{1,0,1}}%
      \expandafter\def\csname LT4\endcsname{\color[rgb]{0,1,1}}%
      \expandafter\def\csname LT5\endcsname{\color[rgb]{1,1,0}}%
      \expandafter\def\csname LT6\endcsname{\color[rgb]{0,0,0}}%
      \expandafter\def\csname LT7\endcsname{\color[rgb]{1,0.3,0}}%
      \expandafter\def\csname LT8\endcsname{\color[rgb]{0.5,0.5,0.5}}%
    \else
      % gray
      \def\colorrgb#1{\color{black}}%
      \def\colorgray#1{\color[gray]{#1}}%
      \expandafter\def\csname LTw\endcsname{\color{white}}%
      \expandafter\def\csname LTb\endcsname{\color{black}}%
      \expandafter\def\csname LTa\endcsname{\color{black}}%
      \expandafter\def\csname LT0\endcsname{\color{black}}%
      \expandafter\def\csname LT1\endcsname{\color{black}}%
      \expandafter\def\csname LT2\endcsname{\color{black}}%
      \expandafter\def\csname LT3\endcsname{\color{black}}%
      \expandafter\def\csname LT4\endcsname{\color{black}}%
      \expandafter\def\csname LT5\endcsname{\color{black}}%
      \expandafter\def\csname LT6\endcsname{\color{black}}%
      \expandafter\def\csname LT7\endcsname{\color{black}}%
      \expandafter\def\csname LT8\endcsname{\color{black}}%
    \fi
  \fi
    \setlength{\unitlength}{0.0500bp}%
    \ifx\gptboxheight\undefined%
      \newlength{\gptboxheight}%
      \newlength{\gptboxwidth}%
      \newsavebox{\gptboxtext}%
    \fi%
    \setlength{\fboxrule}{0.5pt}%
    \setlength{\fboxsep}{1pt}%
\begin{picture}(6800.00,5660.00)%
    \gplgaddtomacro\gplbacktext{%
      \csname LTb\endcsname%
      \put(747,595){\makebox(0,0)[r]{\strut{}$10^{-9}$}}%
      \csname LTb\endcsname%
      \put(747,1203){\makebox(0,0)[r]{\strut{}$10^{-8}$}}%
      \csname LTb\endcsname%
      \put(747,1810){\makebox(0,0)[r]{\strut{}$10^{-7}$}}%
      \csname LTb\endcsname%
      \put(747,2418){\makebox(0,0)[r]{\strut{}$10^{-6}$}}%
      \csname LTb\endcsname%
      \put(747,3025){\makebox(0,0)[r]{\strut{}$10^{-5}$}}%
      \csname LTb\endcsname%
      \put(747,3633){\makebox(0,0)[r]{\strut{}$10^{-4}$}}%
      \csname LTb\endcsname%
      \put(747,4240){\makebox(0,0)[r]{\strut{}$10^{-3}$}}%
      \csname LTb\endcsname%
      \put(747,4848){\makebox(0,0)[r]{\strut{}$10^{-2}$}}%
      \csname LTb\endcsname%
      \put(747,5455){\makebox(0,0)[r]{\strut{}$10^{-1}$}}%
      \csname LTb\endcsname%
      \put(849,409){\makebox(0,0){\strut{}$0.08$}}%
      \csname LTb\endcsname%
      \put(1513,409){\makebox(0,0){\strut{}$0.10$}}%
      \csname LTb\endcsname%
      \put(2177,409){\makebox(0,0){\strut{}$0.12$}}%
      \csname LTb\endcsname%
      \put(2841,409){\makebox(0,0){\strut{}$0.14$}}%
      \csname LTb\endcsname%
      \put(3505,409){\makebox(0,0){\strut{}$0.16$}}%
      \csname LTb\endcsname%
      \put(4169,409){\makebox(0,0){\strut{}$0.18$}}%
      \csname LTb\endcsname%
      \put(4833,409){\makebox(0,0){\strut{}$0.20$}}%
      \csname LTb\endcsname%
      \put(5497,409){\makebox(0,0){\strut{}$0.22$}}%
      \csname LTb\endcsname%
      \put(6161,409){\makebox(0,0){\strut{}$0.24$}}%
    }%
    \gplgaddtomacro\gplfronttext{%
      \csname LTb\endcsname%
      \put(144,3025){\rotatebox{-270}{\makebox(0,0){\strut{}Frame error rate}}}%
      \csname LTb\endcsname%
      \put(3671,130){\makebox(0,0){\strut{}Symbol error probability $(p)$}}%
      \csname LTb\endcsname%
      \put(4555,1599){\makebox(0,0)[l]{\strut{}Iterative}}%
      \csname LTb\endcsname%
      \put(4555,1413){\makebox(0,0)[l]{\strut{}pp -- Kreshchuk}}%
      \csname LTb\endcsname%
      \put(4555,1227){\makebox(0,0)[l]{\strut{}pp -- Emmadi}}%
      \csname LTb\endcsname%
      \put(4555,1041){\makebox(0,0)[l]{\strut{}pp -- mod. Condo}}%
      \csname LTb\endcsname%
      \put(4555,855){\makebox(0,0)[l]{\strut{}pp -- proposed}}%
    }%
    \gplbacktext
    \put(0,0){\includegraphics{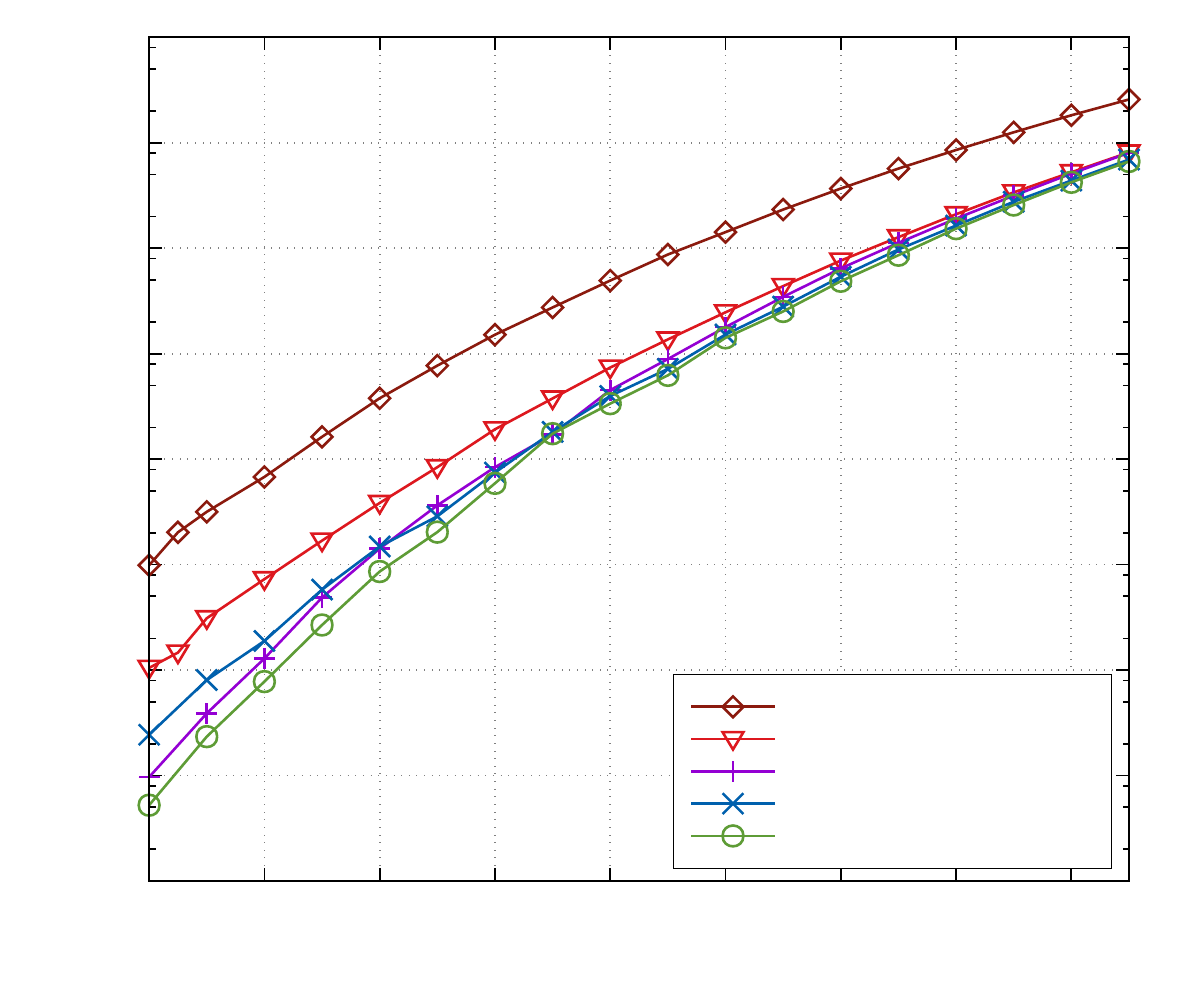}}%
    \gplfronttext
  \end{picture}%
\endgroup

%% file: images/sim256_5-2-4.tex
% GNUPLOT: LaTeX picture with Postscript
\begingroup
  \makeatletter
  \providecommand\color[2][]{%
    \GenericError{(gnuplot) \space\space\space\@spaces}{%
      Package color not loaded in conjunction with
      terminal option `colourtext'%
    }{See the gnuplot documentation for explanation.%
    }{Either use 'blacktext' in gnuplot or load the package
      color.sty in LaTeX.}%
    \renewcommand\color[2][]{}%
  }%
  \providecommand\includegraphics[2][]{%
    \GenericError{(gnuplot) \space\space\space\@spaces}{%
      Package graphicx or graphics not loaded%
    }{See the gnuplot documentation for explanation.%
    }{The gnuplot epslatex terminal needs graphicx.sty or graphics.sty.}%
    \renewcommand\includegraphics[2][]{}%
  }%
  \providecommand\rotatebox[2]{#2}%
  \@ifundefined{ifGPcolor}{%
    \newif\ifGPcolor
    \GPcolortrue
  }{}%
  \@ifundefined{ifGPblacktext}{%
    \newif\ifGPblacktext
    \GPblacktexttrue
  }{}%
  % define a \g@addto@macro without @ in the name:
  \let\gplgaddtomacro\g@addto@macro
  % define empty templates for all commands taking text:
  \gdef\gplbacktext{}%
  \gdef\gplfronttext{}%
  \makeatother
  \ifGPblacktext
    % no textcolor at all
    \def\colorrgb#1{}%
    \def\colorgray#1{}%
  \else
    % gray or color?
    \ifGPcolor
      \def\colorrgb#1{\color[rgb]{#1}}%
      \def\colorgray#1{\color[gray]{#1}}%
      \expandafter\def\csname LTw\endcsname{\color{white}}%
      \expandafter\def\csname LTb\endcsname{\color{black}}%
      \expandafter\def\csname LTa\endcsname{\color{black}}%
      \expandafter\def\csname LT0\endcsname{\color[rgb]{1,0,0}}%
      \expandafter\def\csname LT1\endcsname{\color[rgb]{0,1,0}}%
      \expandafter\def\csname LT2\endcsname{\color[rgb]{0,0,1}}%
      \expandafter\def\csname LT3\endcsname{\color[rgb]{1,0,1}}%
      \expandafter\def\csname LT4\endcsname{\color[rgb]{0,1,1}}%
      \expandafter\def\csname LT5\endcsname{\color[rgb]{1,1,0}}%
      \expandafter\def\csname LT6\endcsname{\color[rgb]{0,0,0}}%
      \expandafter\def\csname LT7\endcsname{\color[rgb]{1,0.3,0}}%
      \expandafter\def\csname LT8\endcsname{\color[rgb]{0.5,0.5,0.5}}%
    \else
      % gray
      \def\colorrgb#1{\color{black}}%
      \def\colorgray#1{\color[gray]{#1}}%
      \expandafter\def\csname LTw\endcsname{\color{white}}%
      \expandafter\def\csname LTb\endcsname{\color{black}}%
      \expandafter\def\csname LTa\endcsname{\color{black}}%
      \expandafter\def\csname LT0\endcsname{\color{black}}%
      \expandafter\def\csname LT1\endcsname{\color{black}}%
      \expandafter\def\csname LT2\endcsname{\color{black}}%
      \expandafter\def\csname LT3\endcsname{\color{black}}%
      \expandafter\def\csname LT4\endcsname{\color{black}}%
      \expandafter\def\csname LT5\endcsname{\color{black}}%
      \expandafter\def\csname LT6\endcsname{\color{black}}%
      \expandafter\def\csname LT7\endcsname{\color{black}}%
      \expandafter\def\csname LT8\endcsname{\color{black}}%
    \fi
  \fi
    \setlength{\unitlength}{0.0500bp}%
    \ifx\gptboxheight\undefined%
      \newlength{\gptboxheight}%
      \newlength{\gptboxwidth}%
      \newsavebox{\gptboxtext}%
    \fi%
    \setlength{\fboxrule}{0.5pt}%
    \setlength{\fboxsep}{1pt}%
\begin{picture}(6800.00,5660.00)%
    \gplgaddtomacro\gplbacktext{%
      \csname LTb\endcsname%
      \put(747,595){\makebox(0,0)[r]{\strut{}$10^{-9}$}}%
      \csname LTb\endcsname%
      \put(747,1135){\makebox(0,0)[r]{\strut{}$10^{-8}$}}%
      \csname LTb\endcsname%
      \put(747,1675){\makebox(0,0)[r]{\strut{}$10^{-7}$}}%
      \csname LTb\endcsname%
      \put(747,2215){\makebox(0,0)[r]{\strut{}$10^{-6}$}}%
      \csname LTb\endcsname%
      \put(747,2755){\makebox(0,0)[r]{\strut{}$10^{-5}$}}%
      \csname LTb\endcsname%
      \put(747,3295){\makebox(0,0)[r]{\strut{}$10^{-4}$}}%
      \csname LTb\endcsname%
      \put(747,3835){\makebox(0,0)[r]{\strut{}$10^{-3}$}}%
      \csname LTb\endcsname%
      \put(747,4375){\makebox(0,0)[r]{\strut{}$10^{-2}$}}%
      \csname LTb\endcsname%
      \put(747,4915){\makebox(0,0)[r]{\strut{}$10^{-1}$}}%
      \csname LTb\endcsname%
      \put(747,5455){\makebox(0,0)[r]{\strut{}$10^{0}$}}%
      \csname LTb\endcsname%
      \put(849,409){\makebox(0,0){\strut{}$0.01$}}%
      \csname LTb\endcsname%
      \put(1476,409){\makebox(0,0){\strut{}$0.02$}}%
      \csname LTb\endcsname%
      \put(2103,409){\makebox(0,0){\strut{}$0.03$}}%
      \csname LTb\endcsname%
      \put(2730,409){\makebox(0,0){\strut{}$0.04$}}%
      \csname LTb\endcsname%
      \put(3357,409){\makebox(0,0){\strut{}$0.05$}}%
      \csname LTb\endcsname%
      \put(3985,409){\makebox(0,0){\strut{}$0.06$}}%
      \csname LTb\endcsname%
      \put(4612,409){\makebox(0,0){\strut{}$0.07$}}%
      \csname LTb\endcsname%
      \put(5239,409){\makebox(0,0){\strut{}$0.08$}}%
      \csname LTb\endcsname%
      \put(5866,409){\makebox(0,0){\strut{}$0.09$}}%
      \csname LTb\endcsname%
      \put(6493,409){\makebox(0,0){\strut{}$0.10$}}%
    }%
    \gplgaddtomacro\gplfronttext{%
      \csname LTb\endcsname%
      \put(144,3025){\rotatebox{-270}{\makebox(0,0){\strut{}Frame error rate}}}%
      \csname LTb\endcsname%
      \put(3671,130){\makebox(0,0){\strut{}Symbol error probability $(p)$}}%
      \csname LTb\endcsname%
      \put(4555,1599){\makebox(0,0)[l]{\strut{}Iterative}}%
      \csname LTb\endcsname%
      \put(4555,1413){\makebox(0,0)[l]{\strut{}pp -- Kreshchuk}}%
      \csname LTb\endcsname%
      \put(4555,1227){\makebox(0,0)[l]{\strut{}pp -- Emmadi}}%
      \csname LTb\endcsname%
      \put(4555,1041){\makebox(0,0)[l]{\strut{}pp -- mod. Condo}}%
      \csname LTb\endcsname%
      \put(4555,855){\makebox(0,0)[l]{\strut{}pp -- proposed}}%
    }%
    \gplbacktext
    \put(0,0){\includegraphics{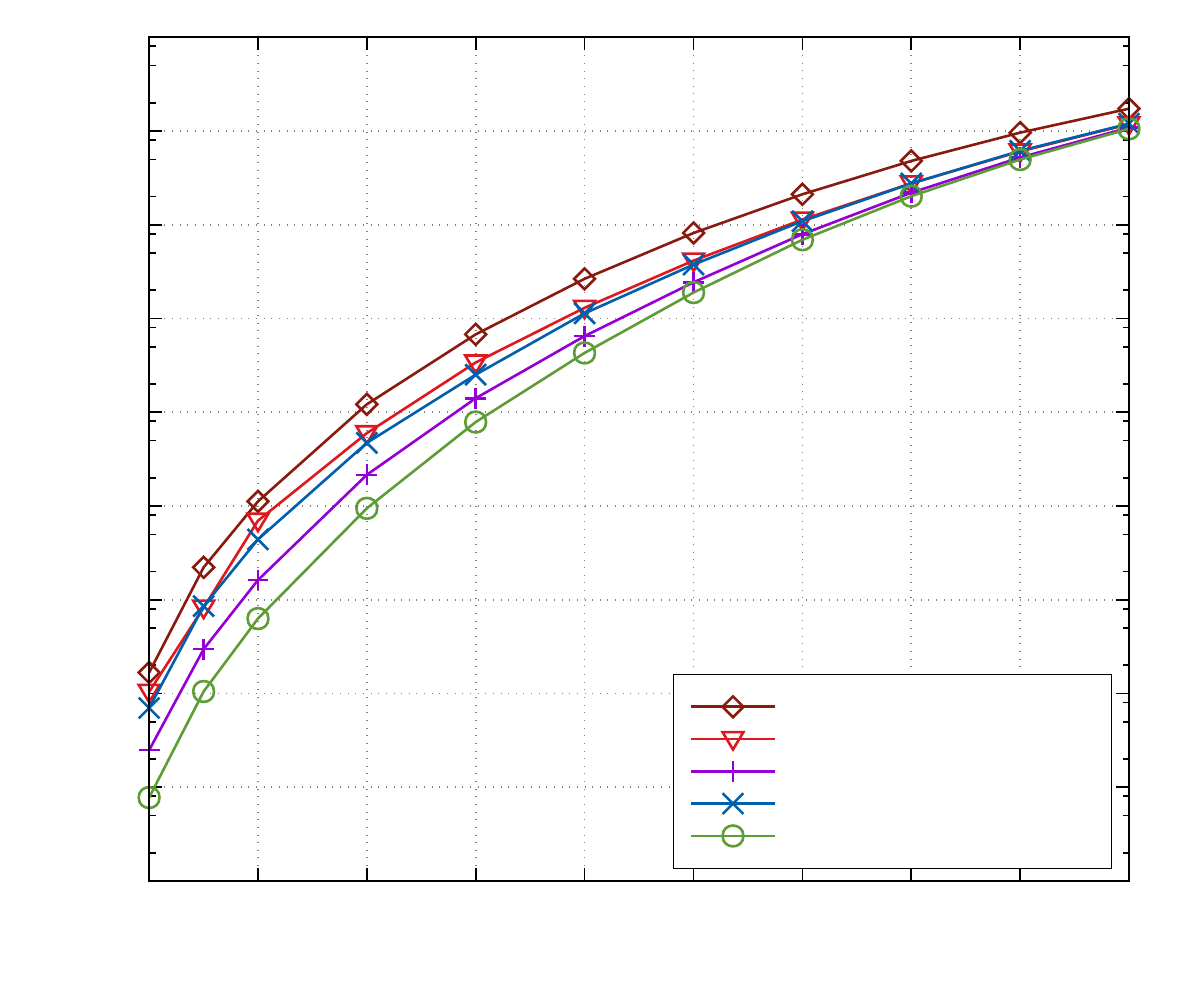}}%
    \gplfronttext
  \end{picture}%
\endgroup

%% file: images/sim256_5-4-4.tex
% GNUPLOT: LaTeX picture with Postscript
\begingroup
  \makeatletter
  \providecommand\color[2][]{%
    \GenericError{(gnuplot) \space\space\space\@spaces}{%
      Package color not loaded in conjunction with
      terminal option `colourtext'%
    }{See the gnuplot documentation for explanation.%
    }{Either use 'blacktext' in gnuplot or load the package
      color.sty in LaTeX.}%
    \renewcommand\color[2][]{}%
  }%
  \providecommand\includegraphics[2][]{%
    \GenericError{(gnuplot) \space\space\space\@spaces}{%
      Package graphicx or graphics not loaded%
    }{See the gnuplot documentation for explanation.%
    }{The gnuplot epslatex terminal needs graphicx.sty or graphics.sty.}%
    \renewcommand\includegraphics[2][]{}%
  }%
  \providecommand\rotatebox[2]{#2}%
  \@ifundefined{ifGPcolor}{%
    \newif\ifGPcolor
    \GPcolortrue
  }{}%
  \@ifundefined{ifGPblacktext}{%
    \newif\ifGPblacktext
    \GPblacktexttrue
  }{}%
  % define a \g@addto@macro without @ in the name:
  \let\gplgaddtomacro\g@addto@macro
  % define empty templates for all commands taking text:
  \gdef\gplbacktext{}%
  \gdef\gplfronttext{}%
  \makeatother
  \ifGPblacktext
    % no textcolor at all
    \def\colorrgb#1{}%
    \def\colorgray#1{}%
  \else
    % gray or color?
    \ifGPcolor
      \def\colorrgb#1{\color[rgb]{#1}}%
      \def\colorgray#1{\color[gray]{#1}}%
      \expandafter\def\csname LTw\endcsname{\color{white}}%
      \expandafter\def\csname LTb\endcsname{\color{black}}%
      \expandafter\def\csname LTa\endcsname{\color{black}}%
      \expandafter\def\csname LT0\endcsname{\color[rgb]{1,0,0}}%
      \expandafter\def\csname LT1\endcsname{\color[rgb]{0,1,0}}%
      \expandafter\def\csname LT2\endcsname{\color[rgb]{0,0,1}}%
      \expandafter\def\csname LT3\endcsname{\color[rgb]{1,0,1}}%
      \expandafter\def\csname LT4\endcsname{\color[rgb]{0,1,1}}%
      \expandafter\def\csname LT5\endcsname{\color[rgb]{1,1,0}}%
      \expandafter\def\csname LT6\endcsname{\color[rgb]{0,0,0}}%
      \expandafter\def\csname LT7\endcsname{\color[rgb]{1,0.3,0}}%
      \expandafter\def\csname LT8\endcsname{\color[rgb]{0.5,0.5,0.5}}%
    \else
      % gray
      \def\colorrgb#1{\color{black}}%
      \def\colorgray#1{\color[gray]{#1}}%
      \expandafter\def\csname LTw\endcsname{\color{white}}%
      \expandafter\def\csname LTb\endcsname{\color{black}}%
      \expandafter\def\csname LTa\endcsname{\color{black}}%
      \expandafter\def\csname LT0\endcsname{\color{black}}%
      \expandafter\def\csname LT1\endcsname{\color{black}}%
      \expandafter\def\csname LT2\endcsname{\color{black}}%
      \expandafter\def\csname LT3\endcsname{\color{black}}%
      \expandafter\def\csname LT4\endcsname{\color{black}}%
      \expandafter\def\csname LT5\endcsname{\color{black}}%
      \expandafter\def\csname LT6\endcsname{\color{black}}%
      \expandafter\def\csname LT7\endcsname{\color{black}}%
      \expandafter\def\csname LT8\endcsname{\color{black}}%
    \fi
  \fi
    \setlength{\unitlength}{0.0500bp}%
    \ifx\gptboxheight\undefined%
      \newlength{\gptboxheight}%
      \newlength{\gptboxwidth}%
      \newsavebox{\gptboxtext}%
    \fi%
    \setlength{\fboxrule}{0.5pt}%
    \setlength{\fboxsep}{1pt}%
\begin{picture}(6800.00,5660.00)%
    \gplgaddtomacro\gplbacktext{%
      \csname LTb\endcsname%
      \put(747,595){\makebox(0,0)[r]{\strut{}$10^{-9}$}}%
      \csname LTb\endcsname%
      \put(747,1203){\makebox(0,0)[r]{\strut{}$10^{-8}$}}%
      \csname LTb\endcsname%
      \put(747,1810){\makebox(0,0)[r]{\strut{}$10^{-7}$}}%
      \csname LTb\endcsname%
      \put(747,2418){\makebox(0,0)[r]{\strut{}$10^{-6}$}}%
      \csname LTb\endcsname%
      \put(747,3025){\makebox(0,0)[r]{\strut{}$10^{-5}$}}%
      \csname LTb\endcsname%
      \put(747,3633){\makebox(0,0)[r]{\strut{}$10^{-4}$}}%
      \csname LTb\endcsname%
      \put(747,4240){\makebox(0,0)[r]{\strut{}$10^{-3}$}}%
      \csname LTb\endcsname%
      \put(747,4848){\makebox(0,0)[r]{\strut{}$10^{-2}$}}%
      \csname LTb\endcsname%
      \put(747,5455){\makebox(0,0)[r]{\strut{}$10^{-1}$}}%
      \csname LTb\endcsname%
      \put(1146,409){\makebox(0,0){\strut{}$0.05$}}%
      \csname LTb\endcsname%
      \put(1740,409){\makebox(0,0){\strut{}$0.06$}}%
      \csname LTb\endcsname%
      \put(2334,409){\makebox(0,0){\strut{}$0.07$}}%
      \csname LTb\endcsname%
      \put(2928,409){\makebox(0,0){\strut{}$0.08$}}%
      \csname LTb\endcsname%
      \put(3522,409){\makebox(0,0){\strut{}$0.09$}}%
      \csname LTb\endcsname%
      \put(4117,409){\makebox(0,0){\strut{}$0.10$}}%
      \csname LTb\endcsname%
      \put(4711,409){\makebox(0,0){\strut{}$0.11$}}%
      \csname LTb\endcsname%
      \put(5305,409){\makebox(0,0){\strut{}$0.12$}}%
      \csname LTb\endcsname%
      \put(5899,409){\makebox(0,0){\strut{}$0.13$}}%
      \csname LTb\endcsname%
      \put(6493,409){\makebox(0,0){\strut{}$0.14$}}%
    }%
    \gplgaddtomacro\gplfronttext{%
      \csname LTb\endcsname%
      \put(144,3025){\rotatebox{-270}{\makebox(0,0){\strut{}Frame error rate}}}%
      \csname LTb\endcsname%
      \put(3671,130){\makebox(0,0){\strut{}Symbol error probability $(p)$}}%
      \csname LTb\endcsname%
      \put(4555,1599){\makebox(0,0)[l]{\strut{}Iterative}}%
      \csname LTb\endcsname%
      \put(4555,1413){\makebox(0,0)[l]{\strut{}pp -- Kreshchuk}}%
      \csname LTb\endcsname%
      \put(4555,1227){\makebox(0,0)[l]{\strut{}pp -- Emmadi}}%
      \csname LTb\endcsname%
      \put(4555,1041){\makebox(0,0)[l]{\strut{}pp -- mod. Condo}}%
      \csname LTb\endcsname%
      \put(4555,855){\makebox(0,0)[l]{\strut{}pp -- proposed}}%
    }%
    \gplbacktext
    \put(0,0){\includegraphics{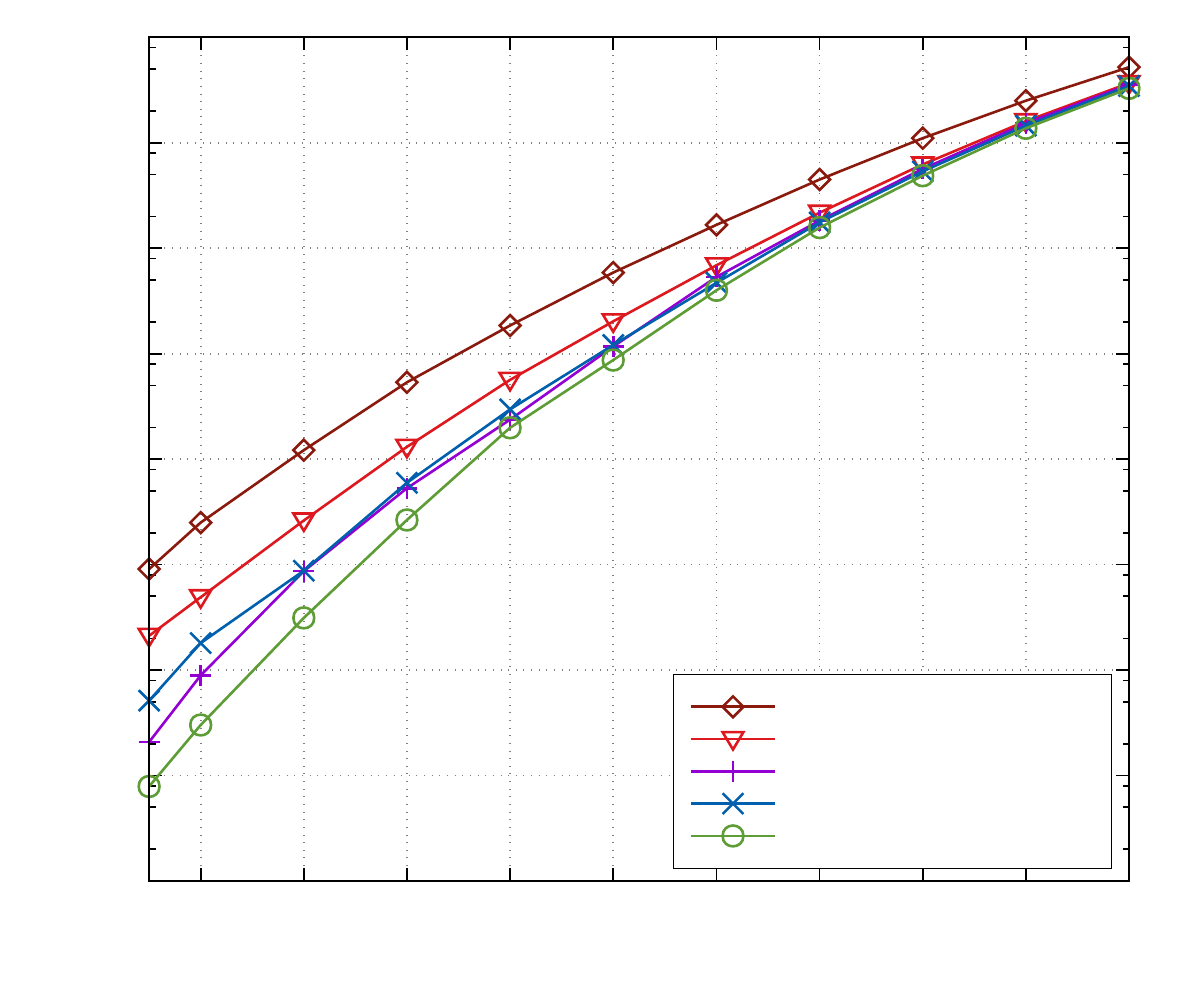}}%
    \gplfronttext
  \end{picture}%
\endgroup

%% file: images/sim1024_8-2-4.tex
% GNUPLOT: LaTeX picture with Postscript
\begingroup
  \makeatletter
  \providecommand\color[2][]{%
    \GenericError{(gnuplot) \space\space\space\@spaces}{%
      Package color not loaded in conjunction with
      terminal option `colourtext'%
    }{See the gnuplot documentation for explanation.%
    }{Either use 'blacktext' in gnuplot or load the package
      color.sty in LaTeX.}%
    \renewcommand\color[2][]{}%
  }%
  \providecommand\includegraphics[2][]{%
    \GenericError{(gnuplot) \space\space\space\@spaces}{%
      Package graphicx or graphics not loaded%
    }{See the gnuplot documentation for explanation.%
    }{The gnuplot epslatex terminal needs graphicx.sty or graphics.sty.}%
    \renewcommand\includegraphics[2][]{}%
  }%
  \providecommand\rotatebox[2]{#2}%
  \@ifundefined{ifGPcolor}{%
    \newif\ifGPcolor
    \GPcolortrue
  }{}%
  \@ifundefined{ifGPblacktext}{%
    \newif\ifGPblacktext
    \GPblacktexttrue
  }{}%
  % define a \g@addto@macro without @ in the name:
  \let\gplgaddtomacro\g@addto@macro
  % define empty templates for all commands taking text:
  \gdef\gplbacktext{}%
  \gdef\gplfronttext{}%
  \makeatother
  \ifGPblacktext
    % no textcolor at all
    \def\colorrgb#1{}%
    \def\colorgray#1{}%
  \else
    % gray or color?
    \ifGPcolor
      \def\colorrgb#1{\color[rgb]{#1}}%
      \def\colorgray#1{\color[gray]{#1}}%
      \expandafter\def\csname LTw\endcsname{\color{white}}%
      \expandafter\def\csname LTb\endcsname{\color{black}}%
      \expandafter\def\csname LTa\endcsname{\color{black}}%
      \expandafter\def\csname LT0\endcsname{\color[rgb]{1,0,0}}%
      \expandafter\def\csname LT1\endcsname{\color[rgb]{0,1,0}}%
      \expandafter\def\csname LT2\endcsname{\color[rgb]{0,0,1}}%
      \expandafter\def\csname LT3\endcsname{\color[rgb]{1,0,1}}%
      \expandafter\def\csname LT4\endcsname{\color[rgb]{0,1,1}}%
      \expandafter\def\csname LT5\endcsname{\color[rgb]{1,1,0}}%
      \expandafter\def\csname LT6\endcsname{\color[rgb]{0,0,0}}%
      \expandafter\def\csname LT7\endcsname{\color[rgb]{1,0.3,0}}%
      \expandafter\def\csname LT8\endcsname{\color[rgb]{0.5,0.5,0.5}}%
    \else
      % gray
      \def\colorrgb#1{\color{black}}%
      \def\colorgray#1{\color[gray]{#1}}%
      \expandafter\def\csname LTw\endcsname{\color{white}}%
      \expandafter\def\csname LTb\endcsname{\color{black}}%
      \expandafter\def\csname LTa\endcsname{\color{black}}%
      \expandafter\def\csname LT0\endcsname{\color{black}}%
      \expandafter\def\csname LT1\endcsname{\color{black}}%
      \expandafter\def\csname LT2\endcsname{\color{black}}%
      \expandafter\def\csname LT3\endcsname{\color{black}}%
      \expandafter\def\csname LT4\endcsname{\color{black}}%
      \expandafter\def\csname LT5\endcsname{\color{black}}%
      \expandafter\def\csname LT6\endcsname{\color{black}}%
      \expandafter\def\csname LT7\endcsname{\color{black}}%
      \expandafter\def\csname LT8\endcsname{\color{black}}%
    \fi
  \fi
    \setlength{\unitlength}{0.0500bp}%
    \ifx\gptboxheight\undefined%
      \newlength{\gptboxheight}%
      \newlength{\gptboxwidth}%
      \newsavebox{\gptboxtext}%
    \fi%
    \setlength{\fboxrule}{0.5pt}%
    \setlength{\fboxsep}{1pt}%
\begin{picture}(6800.00,5660.00)%
    \gplgaddtomacro\gplbacktext{%
      \csname LTb\endcsname%
      \put(747,595){\makebox(0,0)[r]{\strut{}$10^{-9}$}}%
      \csname LTb\endcsname%
      \put(747,1135){\makebox(0,0)[r]{\strut{}$10^{-8}$}}%
      \csname LTb\endcsname%
      \put(747,1675){\makebox(0,0)[r]{\strut{}$10^{-7}$}}%
      \csname LTb\endcsname%
      \put(747,2215){\makebox(0,0)[r]{\strut{}$10^{-6}$}}%
      \csname LTb\endcsname%
      \put(747,2755){\makebox(0,0)[r]{\strut{}$10^{-5}$}}%
      \csname LTb\endcsname%
      \put(747,3295){\makebox(0,0)[r]{\strut{}$10^{-4}$}}%
      \csname LTb\endcsname%
      \put(747,3835){\makebox(0,0)[r]{\strut{}$10^{-3}$}}%
      \csname LTb\endcsname%
      \put(747,4375){\makebox(0,0)[r]{\strut{}$10^{-2}$}}%
      \csname LTb\endcsname%
      \put(747,4915){\makebox(0,0)[r]{\strut{}$10^{-1}$}}%
      \csname LTb\endcsname%
      \put(747,5455){\makebox(0,0)[r]{\strut{}$10^{0}$}}%
      \csname LTb\endcsname%
      \put(849,409){\makebox(0,0){\strut{}$0.01$}}%
      \csname LTb\endcsname%
      \put(1978,409){\makebox(0,0){\strut{}$0.02$}}%
      \csname LTb\endcsname%
      \put(3107,409){\makebox(0,0){\strut{}$0.03$}}%
      \csname LTb\endcsname%
      \put(4235,409){\makebox(0,0){\strut{}$0.04$}}%
      \csname LTb\endcsname%
      \put(5364,409){\makebox(0,0){\strut{}$0.05$}}%
      \csname LTb\endcsname%
      \put(6493,409){\makebox(0,0){\strut{}$0.06$}}%
    }%
    \gplgaddtomacro\gplfronttext{%
      \csname LTb\endcsname%
      \put(144,3025){\rotatebox{-270}{\makebox(0,0){\strut{}Frame error rate}}}%
      \csname LTb\endcsname%
      \put(3671,130){\makebox(0,0){\strut{}Symbol error probability $(p)$}}%
      \csname LTb\endcsname%
      \put(4555,1599){\makebox(0,0)[l]{\strut{}Iterative}}%
      \csname LTb\endcsname%
      \put(4555,1413){\makebox(0,0)[l]{\strut{}pp -- Kreshchuk}}%
      \csname LTb\endcsname%
      \put(4555,1227){\makebox(0,0)[l]{\strut{}pp -- Emmadi}}%
      \csname LTb\endcsname%
      \put(4555,1041){\makebox(0,0)[l]{\strut{}pp -- mod. Condo}}%
      \csname LTb\endcsname%
      \put(4555,855){\makebox(0,0)[l]{\strut{}pp -- proposed}}%
    }%
    \gplbacktext
    \put(0,0){\includegraphics{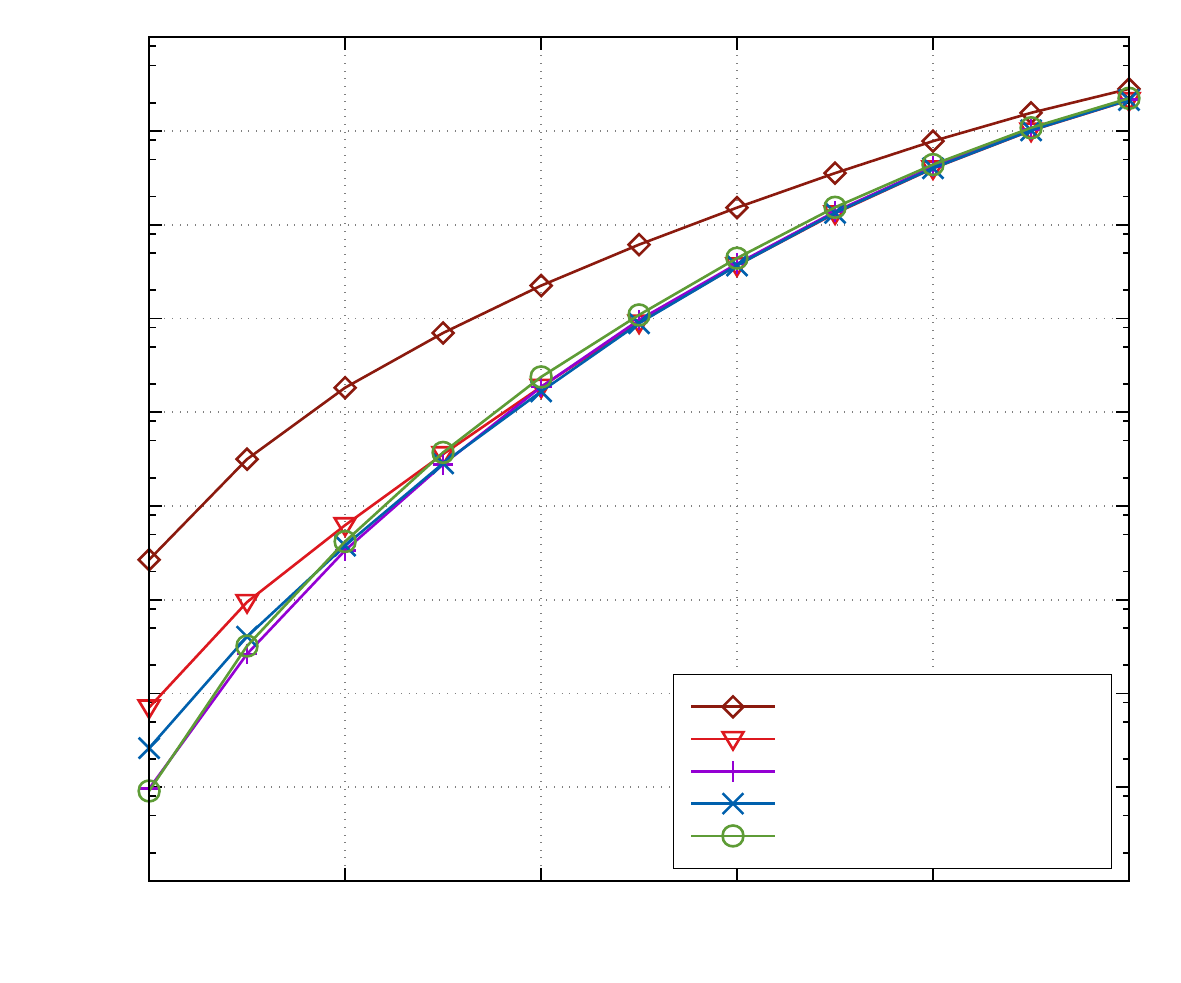}}%
    \gplfronttext
  \end{picture}%
\endgroup

%% file: images/sim2304-2-4.tex
% GNUPLOT: LaTeX picture with Postscript
\begingroup
  \makeatletter
  \providecommand\color[2][]{%
    \GenericError{(gnuplot) \space\space\space\@spaces}{%
      Package color not loaded in conjunction with
      terminal option `colourtext'%
    }{See the gnuplot documentation for explanation.%
    }{Either use 'blacktext' in gnuplot or load the package
      color.sty in LaTeX.}%
    \renewcommand\color[2][]{}%
  }%
  \providecommand\includegraphics[2][]{%
    \GenericError{(gnuplot) \space\space\space\@spaces}{%
      Package graphicx or graphics not loaded%
    }{See the gnuplot documentation for explanation.%
    }{The gnuplot epslatex terminal needs graphicx.sty or graphics.sty.}%
    \renewcommand\includegraphics[2][]{}%
  }%
  \providecommand\rotatebox[2]{#2}%
  \@ifundefined{ifGPcolor}{%
    \newif\ifGPcolor
    \GPcolortrue
  }{}%
  \@ifundefined{ifGPblacktext}{%
    \newif\ifGPblacktext
    \GPblacktexttrue
  }{}%
  % define a \g@addto@macro without @ in the name:
  \let\gplgaddtomacro\g@addto@macro
  % define empty templates for all commands taking text:
  \gdef\gplbacktext{}%
  \gdef\gplfronttext{}%
  \makeatother
  \ifGPblacktext
    % no textcolor at all
    \def\colorrgb#1{}%
    \def\colorgray#1{}%
  \else
    % gray or color?
    \ifGPcolor
      \def\colorrgb#1{\color[rgb]{#1}}%
      \def\colorgray#1{\color[gray]{#1}}%
      \expandafter\def\csname LTw\endcsname{\color{white}}%
      \expandafter\def\csname LTb\endcsname{\color{black}}%
      \expandafter\def\csname LTa\endcsname{\color{black}}%
      \expandafter\def\csname LT0\endcsname{\color[rgb]{1,0,0}}%
      \expandafter\def\csname LT1\endcsname{\color[rgb]{0,1,0}}%
      \expandafter\def\csname LT2\endcsname{\color[rgb]{0,0,1}}%
      \expandafter\def\csname LT3\endcsname{\color[rgb]{1,0,1}}%
      \expandafter\def\csname LT4\endcsname{\color[rgb]{0,1,1}}%
      \expandafter\def\csname LT5\endcsname{\color[rgb]{1,1,0}}%
      \expandafter\def\csname LT6\endcsname{\color[rgb]{0,0,0}}%
      \expandafter\def\csname LT7\endcsname{\color[rgb]{1,0.3,0}}%
      \expandafter\def\csname LT8\endcsname{\color[rgb]{0.5,0.5,0.5}}%
    \else
      % gray
      \def\colorrgb#1{\color{black}}%
      \def\colorgray#1{\color[gray]{#1}}%
      \expandafter\def\csname LTw\endcsname{\color{white}}%
      \expandafter\def\csname LTb\endcsname{\color{black}}%
      \expandafter\def\csname LTa\endcsname{\color{black}}%
      \expandafter\def\csname LT0\endcsname{\color{black}}%
      \expandafter\def\csname LT1\endcsname{\color{black}}%
      \expandafter\def\csname LT2\endcsname{\color{black}}%
      \expandafter\def\csname LT3\endcsname{\color{black}}%
      \expandafter\def\csname LT4\endcsname{\color{black}}%
      \expandafter\def\csname LT5\endcsname{\color{black}}%
      \expandafter\def\csname LT6\endcsname{\color{black}}%
      \expandafter\def\csname LT7\endcsname{\color{black}}%
      \expandafter\def\csname LT8\endcsname{\color{black}}%
    \fi
  \fi
    \setlength{\unitlength}{0.0500bp}%
    \ifx\gptboxheight\undefined%
      \newlength{\gptboxheight}%
      \newlength{\gptboxwidth}%
      \newsavebox{\gptboxtext}%
    \fi%
    \setlength{\fboxrule}{0.5pt}%
    \setlength{\fboxsep}{1pt}%
\begin{picture}(6800.00,5660.00)%
    \gplgaddtomacro\gplbacktext{%
      \csname LTb\endcsname%
      \put(747,595){\makebox(0,0)[r]{\strut{}$10^{-8}$}}%
      \csname LTb\endcsname%
      \put(747,1203){\makebox(0,0)[r]{\strut{}$10^{-7}$}}%
      \csname LTb\endcsname%
      \put(747,1810){\makebox(0,0)[r]{\strut{}$10^{-6}$}}%
      \csname LTb\endcsname%
      \put(747,2418){\makebox(0,0)[r]{\strut{}$10^{-5}$}}%
      \csname LTb\endcsname%
      \put(747,3025){\makebox(0,0)[r]{\strut{}$10^{-4}$}}%
      \csname LTb\endcsname%
      \put(747,3633){\makebox(0,0)[r]{\strut{}$10^{-3}$}}%
      \csname LTb\endcsname%
      \put(747,4240){\makebox(0,0)[r]{\strut{}$10^{-2}$}}%
      \csname LTb\endcsname%
      \put(747,4848){\makebox(0,0)[r]{\strut{}$10^{-1}$}}%
      \csname LTb\endcsname%
      \put(747,5455){\makebox(0,0)[r]{\strut{}$10^{0}$}}%
      \csname LTb\endcsname%
      \put(1283,409){\makebox(0,0){\strut{}$0.01$}}%
      \csname LTb\endcsname%
      \put(3020,409){\makebox(0,0){\strut{}$0.02$}}%
      \csname LTb\endcsname%
      \put(4756,409){\makebox(0,0){\strut{}$0.03$}}%
      \csname LTb\endcsname%
      \put(6493,409){\makebox(0,0){\strut{}$0.04$}}%
    }%
    \gplgaddtomacro\gplfronttext{%
      \csname LTb\endcsname%
      \put(144,3025){\rotatebox{-270}{\makebox(0,0){\strut{}Frame error rate}}}%
      \csname LTb\endcsname%
      \put(3671,130){\makebox(0,0){\strut{}Symbol error probability $(p)$}}%
      \csname LTb\endcsname%
      \put(4555,1599){\makebox(0,0)[l]{\strut{}Iterative}}%
      \csname LTb\endcsname%
      \put(4555,1413){\makebox(0,0)[l]{\strut{}pp -- Kreshchuk}}%
      \csname LTb\endcsname%
      \put(4555,1227){\makebox(0,0)[l]{\strut{}pp -- Emmadi}}%
      \csname LTb\endcsname%
      \put(4555,1041){\makebox(0,0)[l]{\strut{}pp -- mod. Condo}}%
      \csname LTb\endcsname%
      \put(4555,855){\makebox(0,0)[l]{\strut{}pp -- proposed}}%
    }%
    \gplbacktext
    \put(0,0){\includegraphics{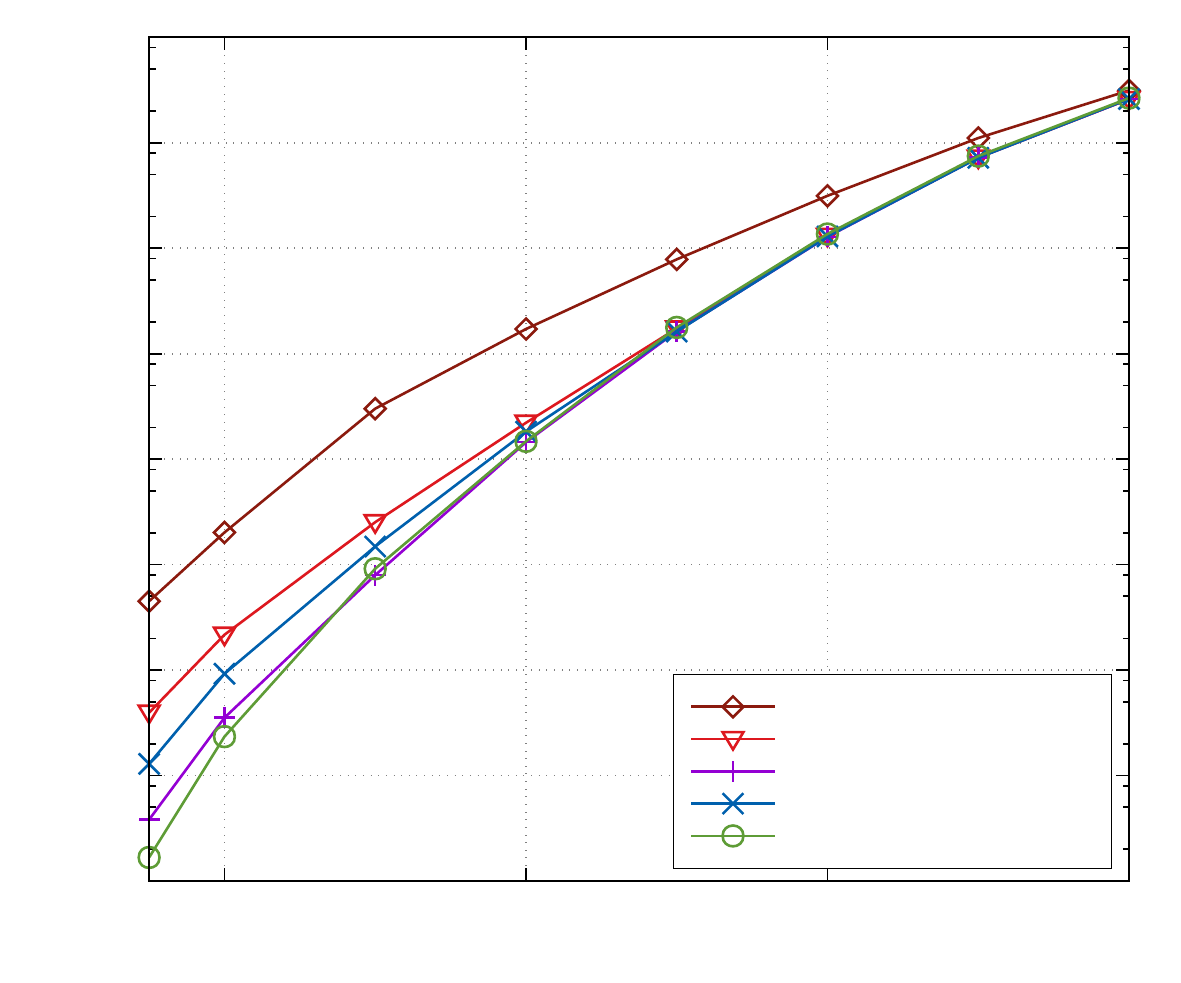}}%
    \gplfronttext
  \end{picture}%
\endgroup

%% file: images/gamma.tex
% GNUPLOT: LaTeX picture with Postscript
\begingroup
  \makeatletter
  \providecommand\color[2][]{%
    \GenericError{(gnuplot) \space\space\space\@spaces}{%
      Package color not loaded in conjunction with
      terminal option `colourtext'%
    }{See the gnuplot documentation for explanation.%
    }{Either use 'blacktext' in gnuplot or load the package
      color.sty in LaTeX.}%
    \renewcommand\color[2][]{}%
  }%
  \providecommand\includegraphics[2][]{%
    \GenericError{(gnuplot) \space\space\space\@spaces}{%
      Package graphicx or graphics not loaded%
    }{See the gnuplot documentation for explanation.%
    }{The gnuplot epslatex terminal needs graphicx.sty or graphics.sty.}%
    \renewcommand\includegraphics[2][]{}%
  }%
  \providecommand\rotatebox[2]{#2}%
  \@ifundefined{ifGPcolor}{%
    \newif\ifGPcolor
    \GPcolortrue
  }{}%
  \@ifundefined{ifGPblacktext}{%
    \newif\ifGPblacktext
    \GPblacktexttrue
  }{}%
  % define a \g@addto@macro without @ in the name:
  \let\gplgaddtomacro\g@addto@macro
  % define empty templates for all commands taking text:
  \gdef\gplbacktext{}%
  \gdef\gplfronttext{}%
  \makeatother
  \ifGPblacktext
    % no textcolor at all
    \def\colorrgb#1{}%
    \def\colorgray#1{}%
  \else
    % gray or color?
    \ifGPcolor
      \def\colorrgb#1{\color[rgb]{#1}}%
      \def\colorgray#1{\color[gray]{#1}}%
      \expandafter\def\csname LTw\endcsname{\color{white}}%
      \expandafter\def\csname LTb\endcsname{\color{black}}%
      \expandafter\def\csname LTa\endcsname{\color{black}}%
      \expandafter\def\csname LT0\endcsname{\color[rgb]{1,0,0}}%
      \expandafter\def\csname LT1\endcsname{\color[rgb]{0,1,0}}%
      \expandafter\def\csname LT2\endcsname{\color[rgb]{0,0,1}}%
      \expandafter\def\csname LT3\endcsname{\color[rgb]{1,0,1}}%
      \expandafter\def\csname LT4\endcsname{\color[rgb]{0,1,1}}%
      \expandafter\def\csname LT5\endcsname{\color[rgb]{1,1,0}}%
      \expandafter\def\csname LT6\endcsname{\color[rgb]{0,0,0}}%
      \expandafter\def\csname LT7\endcsname{\color[rgb]{1,0.3,0}}%
      \expandafter\def\csname LT8\endcsname{\color[rgb]{0.5,0.5,0.5}}%
    \else
      % gray
      \def\colorrgb#1{\color{black}}%
      \def\colorgray#1{\color[gray]{#1}}%
      \expandafter\def\csname LTw\endcsname{\color{white}}%
      \expandafter\def\csname LTb\endcsname{\color{black}}%
      \expandafter\def\csname LTa\endcsname{\color{black}}%
      \expandafter\def\csname LT0\endcsname{\color{black}}%
      \expandafter\def\csname LT1\endcsname{\color{black}}%
      \expandafter\def\csname LT2\endcsname{\color{black}}%
      \expandafter\def\csname LT3\endcsname{\color{black}}%
      \expandafter\def\csname LT4\endcsname{\color{black}}%
      \expandafter\def\csname LT5\endcsname{\color{black}}%
      \expandafter\def\csname LT6\endcsname{\color{black}}%
      \expandafter\def\csname LT7\endcsname{\color{black}}%
      \expandafter\def\csname LT8\endcsname{\color{black}}%
    \fi
  \fi
    \setlength{\unitlength}{0.0500bp}%
    \ifx\gptboxheight\undefined%
      \newlength{\gptboxheight}%
      \newlength{\gptboxwidth}%
      \newsavebox{\gptboxtext}%
    \fi%
    \setlength{\fboxrule}{0.5pt}%
    \setlength{\fboxsep}{1pt}%
\begin{picture}(6800.00,5660.00)%
    \gplgaddtomacro\gplbacktext{%
      \csname LTb\endcsname%
      \put(747,595){\makebox(0,0)[r]{\strut{}$10^{-7}$}}%
      \csname LTb\endcsname%
      \put(747,1289){\makebox(0,0)[r]{\strut{}$10^{-6}$}}%
      \csname LTb\endcsname%
      \put(747,1984){\makebox(0,0)[r]{\strut{}$10^{-5}$}}%
      \csname LTb\endcsname%
      \put(747,2678){\makebox(0,0)[r]{\strut{}$10^{-4}$}}%
      \csname LTb\endcsname%
      \put(747,3372){\makebox(0,0)[r]{\strut{}$10^{-3}$}}%
      \csname LTb\endcsname%
      \put(747,4066){\makebox(0,0)[r]{\strut{}$10^{-2}$}}%
      \csname LTb\endcsname%
      \put(747,4761){\makebox(0,0)[r]{\strut{}$10^{-1}$}}%
      \csname LTb\endcsname%
      \put(747,5455){\makebox(0,0)[r]{\strut{}$10^{0}$}}%
      \csname LTb\endcsname%
      \put(849,409){\makebox(0,0){\strut{}$10^{-9}$}}%
      \csname LTb\endcsname%
      \put(1476,409){\makebox(0,0){\strut{}$10^{-8}$}}%
      \csname LTb\endcsname%
      \put(2103,409){\makebox(0,0){\strut{}$10^{-7}$}}%
      \csname LTb\endcsname%
      \put(2730,409){\makebox(0,0){\strut{}$10^{-6}$}}%
      \csname LTb\endcsname%
      \put(3357,409){\makebox(0,0){\strut{}$10^{-5}$}}%
      \csname LTb\endcsname%
      \put(3985,409){\makebox(0,0){\strut{}$10^{-4}$}}%
      \csname LTb\endcsname%
      \put(4612,409){\makebox(0,0){\strut{}$10^{-3}$}}%
      \csname LTb\endcsname%
      \put(5239,409){\makebox(0,0){\strut{}$10^{-2}$}}%
      \csname LTb\endcsname%
      \put(5866,409){\makebox(0,0){\strut{}$10^{-1}$}}%
      \csname LTb\endcsname%
      \put(6493,409){\makebox(0,0){\strut{}$10^{0}$}}%
    }%
    \gplgaddtomacro\gplfronttext{%
      \csname LTb\endcsname%
      \put(144,3025){\rotatebox{-270}{\makebox(0,0){\strut{}$\gamma_{\text{{\tt eras}}}$}}}%
      \csname LTb\endcsname%
      \put(3671,130){\makebox(0,0){\strut{}Frame error rate}}%
      \csname LTb\endcsname%
      \put(1637,5102){\makebox(0,0)[l]{\strut{}$[2304,2024,15]_{256}$}}%
      \csname LTb\endcsname%
      \put(1637,4823){\makebox(0,0)[l]{\strut{}$[1024,840,15]_{256}$}}%
      \csname LTb\endcsname%
      \put(1637,4544){\makebox(0,0)[l]{\strut{}$[256,168,15]_{16}$}}%
      \csname LTb\endcsname%
      \put(1637,4265){\makebox(0,0)[l]{\strut{}$[256,144,25]_{16}$}}%
      \csname LTb\endcsname%
      \put(1637,3986){\makebox(0,0)[l]{\strut{}$[64,24,15]_{16}$}}%
      \csname LTb\endcsname%
      \put(1637,3707){\makebox(0,0)[l]{\strut{}$[64,16,25]_{16}$}}%
    }%
    \gplbacktext
    \put(0,0){\includegraphics{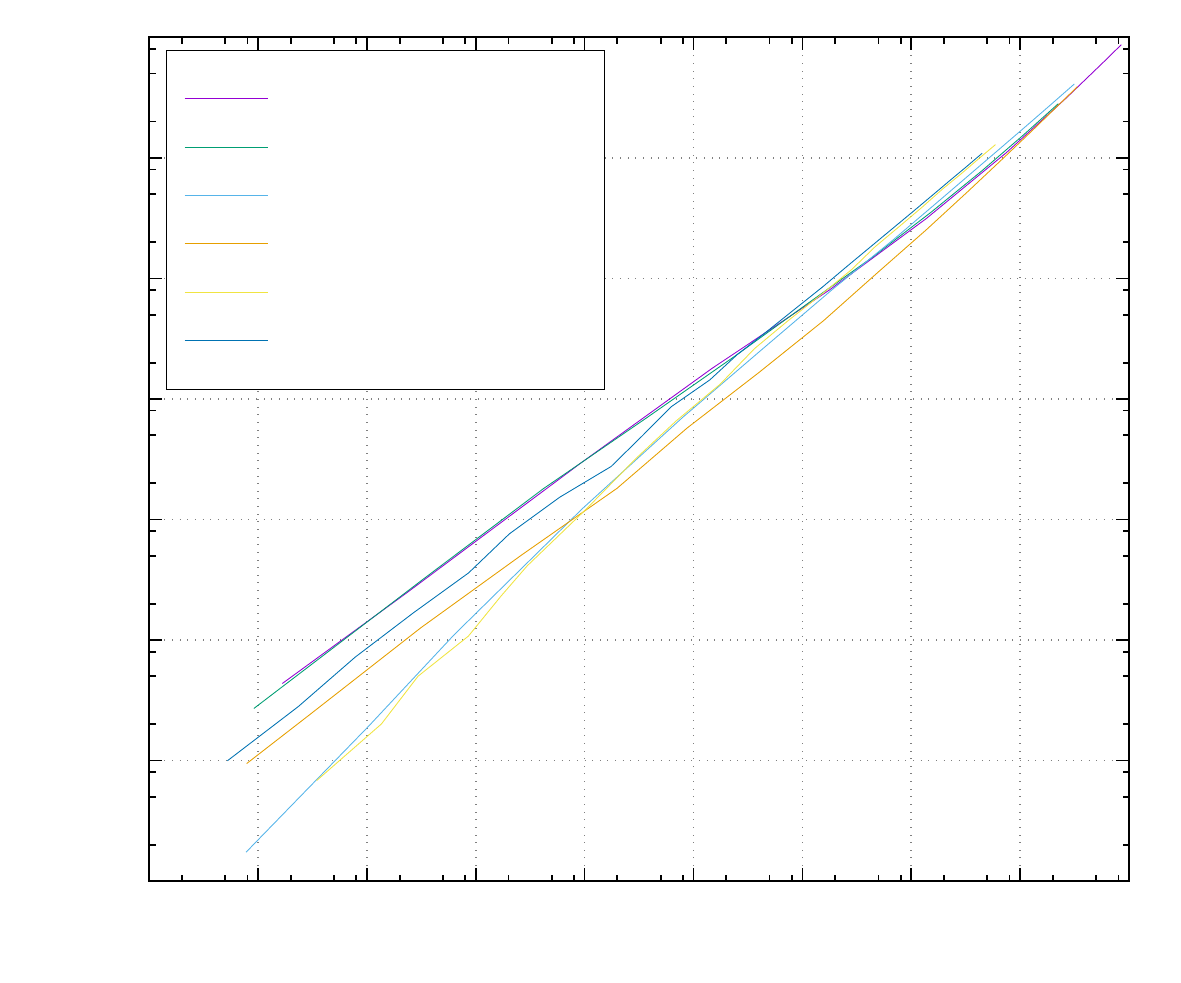}}%
    \gplfronttext
  \end{picture}%
\endgroup